\documentclass[10pt,twocolumn]{article}

\usepackage[
top=1.78cm,
bottom=1.78cm,
left=1.65cm,
right=1.65cm,
headsep=0cm,
]{geometry}
\usepackage{inatel}
\usepackage{times}
\usepackage{enumitem}
\usepackage{graphicx}
\usepackage{url,hyperref}
\usepackage[utf8]{inputenc}
\usepackage{float}
\usepackage{caption}
\usepackage{mathtools}
\usepackage[hang,flushmargin]{footmisc} 
\usepackage{xcolor}
\usepackage{wrapfig} 
\usepackage{fancyhdr}
\usepackage{etoolbox}
\usepackage{adjustbox}
\usepackage{comment}
\usepackage{relsize} 

\usepackage[
    style=numeric,
    sorting=none,
    maxbibnames=10]{biblatex}
\usepackage[english]{babel}
\begin{document}

\title{Evolution, Future of AI, and Singularity}

\author{\large 
Zeki Doruk Erden }

\maketitle

\paragraph{Abstract} \textit{This article critically examines the foundational principles of contemporary AI methods, exploring the limitations that hinder its potential. We draw parallels between the modern AI landscape and the 20\textsuperscript{th}-century Modern Synthesis in evolutionary biology, and highlight how advancements in evolutionary theory that augmented the Modern Synthesis, particularly those of Evolutionary Developmental Biology, offer insights that can inform a new design paradigm for AI. By synthesizing findings across AI and evolutionary theory, we propose a pathway to overcome existing limitations, enabling AI to achieve its aspirational goals. We also examine how this perspective transforms the idea of an AI-driven technological singularity from speculative futurism into a grounded prospect.}

\section{Introduction}
\label{sec:introduction}

Artificial intelligence has steadily integrated into daily life over the past decade \cite{precedenceresearchArtificialIntelligence, mcelheran2024ai}, with finally a direct impact on the life of the average person through the advent of large language models \cite{springsappsLargeLanguage}. What was once deemed speculative science fiction is now becoming an undeniable reality. The excitement surrounding its applications is evident, with industries, governments, and individuals preparing for an AI-driven future, aiming to mitigate risks and seize opportunities. Numerous perspectives on AI's future emerge daily, alongside thousands of research papers aimed at enhancing machine learning systems to advance the field \cite{stanfordIndexReport}. But what lies beneath this hype? What are the foundational principles driving contemporary AI methods, and what assumptions and capabilities do they entail? Can we classify AI as a "solved problem," where the only challenge remaining is the refinement of existing algorithms' numerical performance or their integration to high-level functionalities? Are the current AI technologies equipped to fulfill the expectations they have generated? Or is the future of AI on the brink of a paradigm shift \cite{kuhn1997structure}, necessitating new methods, approaches, assumptions, and capabilities altogether? Furthermore, what connections can we draw between these questions and the recent advancements in evolutionary theory - a field that, at first glance, seems only tangentially related to artificial intelligence?

This article aims to address these questions. In Section \ref{sec:contemporary_ai}, we will present an overview of contemporary AI technologies from a conceptual perspective, analyzing the strengths that underpin both the success and the shortcomings of the machine learning paradigm of the past decade. In Section \ref{sec:evolution}, we will turn our attention to the dominant view of evolutionary biology from the 20\textsuperscript{th} century, highlighting notable parallels with contemporary machine learning—specifically, the explanatory limitations of the former and the capability shortcomings of the latter. We will delve into significant advancements in evolutionary theory over the past few decades that have enhanced this traditional perspective, particularly in ways that are relevant to today’s AI landscape. Section \ref{sec:evolution_intelligence} will hint at an even deeper connection between evolution and artificial intelligence, while Section \ref{sec:ai_future} will leverage these insights to outline what the future of AI techniques should encompass to transcend the limitations of current systems and fully realize the potential we currently associate with speculative futurist scenarios. Finally, in Section \ref{sec:singularity}, we will examine one of the most compelling of these scenarios: the prospect of technological singularity. We will discuss how the concepts presented throughout the article specifically relate to the potential for AI-driven technological acceleration, which may yield transformative effects for the human condition.

\section{Inherent Limits of the Current Machine Learning Paradigm}
\label{sec:contemporary_ai}

\subsection{An overview of contemporary AI}
\label{sec:what_ai_can_do}

AI of the past 10-15 years driven by advances in machine learning (ML) has excelled in many tasks that has been outside the capabilities of classical algorithms of computer science, including the methods like planning or computer vision that have been traditionally regarded among “AI methods” and formed the previous dominant paradigm before the advent of modern machine learning methods. We begin with an overview of contemporary AI, highlighting some key properties, explaining what these systems can achieve, and the factors driving their success.

Modern machine learning, is based primarily on the approach of function approximation. It treats the desired output of the system as a function that is to be approximated (“learned”). The dominant approach in modern AI in this context is called \textit{deep learning} \cite{bengio2017deep}, using a class of architectures called \textit{artificial neural networks} \cite{gurney2018introduction}. A neural network itself is formed of a large number (often ranging from thousands to billions) of simple, homogenous subcomponents, termed \textit{artificial neurons} (with the questionable supposition of a reasonable analogy with neurons in the brain \cite{katyal2021connectionismcomplexitylivingsystems, iyer2022avoiding}) and the connections between them. Each of these components are equipped with continuous, tuneable parameters that are adjusted slowly in a way that the approximation error with the output function is minimized through gradient descent. The universal approximation theorems for neural networks \cite{cybenko1989approximation} assert that any function can be approximated by a neural network, provided it has a sufficiently large number of parameters. However, in practice, what constitutes a “sufficiently large number of parameters” often means “heavily overparameterized networks” \cite{du2018power}, leading to a situation where we end up with significantly more parameters than theoretically necessary to represent the function . This excess is due to the fact that the straightforward gradient descent method requires the network to have ample degrees of freedom to effectively navigate the solution landscape: When the network lands on a plateau that is far from the global optimum, having these additional parameters allows it to adjust itself and break free from the plateau in pursuit of better solutions.

Methods from the previous design paradigm of AI, often referred to as “classical” or “symbolic AI”, frequently struggled to tackle complex, high-dimensional tasks such as language processing, analyzing intricate images, or managing behavior in highly unpredictable environments. In the past decades, large-scale neural networks have become practical to train, primarily due to advancements in computing hardware \cite{khan2021advancements, baji2017gpu} and the widespread availability of data in specific domains \cite{duan2019artificial}. As a result, many tasks previously unsolvable with classical methods—such as advanced computer vision \cite{khan2021machine, chai2021deep}, language processing \cite{zhao2023survey, min2023recent}, and complex behavior \cite{li2017deep}—have become feasible to learn using neural networks, provided there is sufficient training data available.

As AI steadily conquers challenges once thought insurmountable, growing attention is being drawn to fundamental limitations in the principles underlying current algorithms and their ability to integrate into broader systems \cite{clune2019ai,zador2019critique,marcus2018deep,lecun2022path}. What are these limitations that still prevent AI agents from independently operating in the real world, solving complex problems with superior intelligence—beyond the confines of controlled settings like language models or computer vision systems that lack real-world agency? More broadly, what hinders the realization of an AI that continually evolves, wielding the potential for both extraordinary creation and immense destruction? In other words, why does a gap persist between the ambitious visions of early AI futurists and science-fiction dreamers and the reality we face today—despite modern methods being capable of learning virtually any function and solving tasks once believed to be uniquely human just 15 years ago?

\subsection{AI obliterates existing knowledge when learning something new}
\label{sec:ai_destructive_adaptation}

The first major limitation that comes to mind when discussing contemporary AI is its persistent reliance on enormous amounts of data \cite{zha2025data, insideainewsAIsDependency}, a characteristic that has been associated with modern methods since their popularization in the early 2010s. The “function approximation” approach to learning complex functions through intricate, overparameterized networks necessitates that the function being learned is general enough to predict all reasonably expected examples. This can only be achieved if the network is exposed to a vast number of samples, allowing it to grasp the overall patterns and features rather than merely memorizing specific instances. For example, when training a network to recognize dogs, we expect it to identify the common characteristics that define a “dog” instead of focusing on the peculiarities of a few breeds, such as fur color or size, which are not universally applicable to all dogs.

However, this argument is incomplete. It is natural for an AI system to require exposure to a sufficient number of instances to capture the commonalities among them and learn a function that can perform well across the board. If your system can, indeed, learn that golden color or human-scale size are not universal properties of all dogs based solely on images of a few golden retrievers, it’s more probable that your algorithm has a severe bug rather than being an ideal learning algorithm.

The essence of this data-dependency argument highlights something more nuanced: the inefficiency of modern AI systems in utilizing the data available to them—specifically, their inadequacy in effectively integrating new data with what they have already learned. This capability—integrating new knowledge with pre-existing information, often termed \textit{continual learning} in AI literature—is a fundamental expectation from any learning system and is undeniably seen in humans and many other animals, as it could be considered as necessary in the very definition of learning itself. Contemporary AI methods, however, are completely incapable of doing so, as they completely obliterate existing knowledge when learning something new \cite{vandeven2024continuallearningcatastrophicforgetting, hadsell2020embracing}.

\begin{figure}
    \centering
    \includegraphics[width=\linewidth]{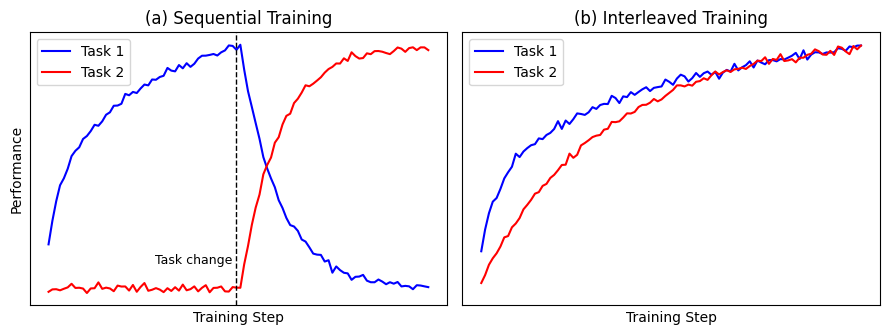}
    \caption{Illustration of the "continual learning" problem, recreated from \cite{vandeven2024continuallearningcatastrophicforgetting}. (a) Sequential training in multiple tasks or environments results the neural network to destroy knowledge of the previous ones, visible as a sharp decrease in performance, a phenomenon frequently called "catastophic forgetting". (b) This is not the case if the network learns multiple tasks in parallel or "replayed" the data pertaining to the previous task when learning a new one, which shows that destruction of past knowledge is not due to any inherent capacity limit or an incommensurability of the tasks/environments.}
    \label{fig:destructive_adaptation}
\end{figure}

Let’s review the mechanism behind this: As discussed earlier, a neural network learns by approximating a function between inputs and outputs, with the target function effectively capturing the net or average effect of a large dataset. By design, this means the function is well-defined and aligned with desired outputs only when all the training data is considered simultaneously. In large, overparameterized networks, this leads to impressive performance—as long as all relevant data is available upfront for a given task (or multiple tasks learned together). However, after this one phase of learning, when a new task (or a new set of tasks—or even the same task under different constraints or in a new environment) is introduced, the function to be learned does not change to be the combination of both the past task and the new task. Instead, the new target function corresponds only to the new task, without any regard to whether the performance on the previous task is retained or not. Experimentally, this often leads to neural networks trained via gradient descent (or similar methods) completely destroying previously learned knowledge when exposed to new training data, effectively erasing prior capabilities (Figure \ref{fig:destructive_adaptation}).\footnote{This phenomenon is often referred to as "catastrophic forgetting" in AI literature. However, this term is misleading, as it implies a gradual decay of unused knowledge—whereas the actual process involves the active destruction of existing knowledge when the system learns a new task. To avoid this mischaracterization, we refrain from using the term and instead describe the process explicitly as what it is (destruction of existing knowledge) or, more concisely, as "destructive adaptation," a term that more accurately captures the phenomenon.}

There are \textit{ad-hoc} fixes for this. The most straightforward, known as replay, involves storing past examples and reintroducing them during new training sessions \cite{rolnick2019experience, buzzega2021rethinking, buzzega2020dark, galashov2023continually}. While simple, this doesn’t truly solve the problem—it merely circumvents it in a way that becomes impractical over time: The longer a system learns, the more past data it must retain and process, making both memory and computational demands explode. Eventually, either new knowledge or older, task-specific knowledge will be diluted to the point of being effectively lost. This approach might be feasible in controlled experimental settings, but it will be highly impractical for real-world agents, which can be expected to encounter millions of distinct situations throughout their lifetime. Other approaches attempt to mitigate this at a higher level but rely on restrictive assumptions—such as clear task boundaries \cite{jacobson2022task, kirkpatrick2017overcoming, wang2022continual, iyer2022avoiding}, external signals indicating the active task \cite{rusu2016progressive}, access to task-relevant data at once as a bulk \cite{erden2024directed, lee2020neural, iyer2022avoiding}, or applying only to tasks with high orthogonality among them \cite{iyer2022avoiding,damleactive}, which clearly do not correspond to the natural flow of learning and experience. All these assumptions fundamentally clash with the needs of real-world learning, where an agent encounters information sequentially, integrating new knowledge with past experiences without the need to store all past data, periodically relive its entire learning history, or rely on neatly defined 'tasks' provided by the world.\footnote{While large language models (LLMs) like ChatGPT may seem capable of performing tasks they weren’t explicitly trained on, they do not overcome the continual learning limitation. Instead, they rely on vast training datasets to generalize across tasks, probabilistically infer responses, and employ an approach called in-context learning \cite{dong2022survey} to adapt temporarily without modifying their internal parameters. (The term “learning” in in-context learning is thus a misnomer if we consider learning to necessitate an actual, permanent update of the system, as is commonly perceived.) Their ability to handle novel or niche tasks stems from pre-existing patterns in their training data, not genuine task-specific learning or rapid adaptability of the model. This distinction matters because, in the case of LLMs, the vast amount of training data required to achieve this capability is \textit{readily available} and can be utilized during the training process as a whole, represented by the \textit{internet}, a massive corpus of human language that has been accumulated over decades in a decentralized, global manner. Crucially, language models do not require explicit labels, as their primary objective is to predict the next word, making this data exceptionally well-suited for their training paradigm. Could, for example, a “large behavior model” trained for real-world interaction achieve similar capabilities? This is possible in principle—if we had an equivalent of an internet-scale dataset composed of real-world interactions. But is such a dataset reasonable to expect in the near future? That remains highly questionable. And even if such a corpus could be constructed, one fundamental limitation would still remain: no model, no matter how large, can learn from the sort of data that simply \textit{does not exist} in its training set. This constraint may not be critical for AI systems performing routine tasks that humans already do, but it becomes a significant barrier to AI’s true potential—expanding knowledge itself and dynamically upgrading its understanding as a result of this process (Section \ref{sec:singularity}).}

Why is “continual learning” crucial? At its core, it’s essential for acquiring any new knowledge. More specifically, in this context, it plays a vital role in enabling adaptability to different environments: when a specific task must be executed across varied settings (for instance, applying a learned cooking skill in different kitchens), the system should be able to absorb the nuances of the new environment (such as the location of utensils, available ingredients, and the kitchen layout) without losing sight of the task’s overall principles. Practically, this consideration is significant in many scenarios—especially when the agent’s operational environment or tasks are somewhat known or can be anticipated, but not entirely. Two key examples of this are simulation-to-reality transfer (which involves adapting to changes in sensory fidelity) \cite{zhao2020sim} and situations where an agent must perform tasks that may not be known during training (like household or assistant robots, whose responsibilities are tailored to individual users and cannot be fully predicted in advance) \cite{kejriwal2024challenges, tu2024towards, huang2019neural}. While the experience required to establish a foundation for such agents—grounded in common scenarios and a fundamental understanding of the world and core tasks—may be extensive, this data is often inexpensive and readily available (whether in the form of textual or visual data from the internet or through simulated exposure to the external world). As long as rapid adaptability is possible without sacrificing existing knowledge, the data requirements of AI becomes a manageable issue. To reiterate the central point from the previous paragraphs: The challenge with contemporary AI's data requirements arises not from an absolute scarcity of data, but from inefficiencies in its utilization. This issue is primarily tied to the inability to learn "continually," which prevents the integration of new knowledge into the model without compromising existing knowledge.

\subsection{AI is incomprehensible and non-engineerable}
\label{sec:ai_incomprehensible}

Yet, this is not the only significant limitation of contemporary AI. Another crucial limitation, often overlooked yet potentially more impactful, could serve as a major barrier to its widespread adoption for critical tasks or even pose an existential risk for humanity. This issue revolves around the \textit{total incomprehensibility} of the internal structure of AI systems \cite{marcus2018deep}.

As discussed earlier, the success of a neural network's learning process is intricately linked to its heavy overparameterization. Additionally, neural networks derive their representation from continuous parameters, often fine-tuned to a few significant digits, and exhibit a nonlinear response dependent on hundreds or thousands of inputs \textit{per neuron}. All of this occurs within a flat network devoid of any internal structure, composed of extensively connected units, aside from a few biases related to the observation space (e.g., shared parameters in architectures like convolutional neural networks \cite{o2015introduction}) in certain applications, which neither enhance comprehensibility nor designed to do so in the first place. There is currently no mainstream method that can reliably incorporate a modular and hierarchical structure into neural networks to enhance their comprehensibility (with existing attempts \cite{su2024focuslearn, goyal2020object, pateria2021hierarchical} not addressing the inherent incomprehensibility of their foundational components, which remain overparameterized, continuous, and nonlinear neural networks at the lowest levels). Research aimed at explaining their internal operations primarily focuses on \textit{post-hoc} analyses of responses and learned parameter statistics, rather than tackling the intrinsic incomprehensibility of the learned systems themselves \cite{xu2019explainable}.

This stands in stark contrast to nearly all engineering systems created by humans to date. It is surely difficult to expect a system that has learned from a huge amount of data to be representing all this knowledge with a small number of parameters, or to be immediately comprehensible at every level - yet it should be reasonable to expect (or desire, since AI is ultimately an engineering system designed for our use) some level of comprehensibility at a higher level of abstraction, or at each level of organization when examined in isolation. This lack of comprehensibility not only hampers our understanding of what these systems do, but also it inhibits decomposability and low-level modifiability at the internal component level, rendering AI inherently uncontrollable and unreliable, aside from its statistical guarantees.

Additionally, and still very importantly, there is a closely related limitation of current machine learning systems: their incomprehensibility and non-decomposability impede the integration of the internal representations they learn with human-designed algorithms or established knowledge bases. This encompasses not only explicit, already-known information about the agent's operational domain (for example, enabling an agent to be pre-informed of a existing map of an environment instead of requiring extensive interaction to learn the environment) but also a broad range of methods under the umbrella of classical or symbolic AI, such as search, logic, planning, and constraint satisfaction \cite{norvig2002modern}. This integration is highly desirable, as symbolic AI—and the field of computer science more broadly—boasts decades of research and robust, well-developed methods that excel in tasks that cannot be efficiently accomplished through data alone. These methods possess highly sought-after properties, including precision, reliability, and inherent comprehensibility. However, the current machine learning paradigm, by its very nature, is unable to harness this potential.

An illustrative example of this challenge, as well as one of the most prominent fields that offers significant potential, is deliberative behavior and planning. Deliberative behavior allows an AI agent to make precise, goal-driven behavioral decisions while avoiding undesired outcomes using an environment model, without needing any reinteraction with the external world to learn desirable behavior. The field is supported by extensive planning literature \cite{ghallab2016automated} stretching to the earliest days of computers. Despite extensive research within the current machine learning paradigm, a mainstream, effective method for integrating deliberative processes with learning systems—particularly one that executes this process in a targeted, goal-oriented manner similar to conventional planning rather than relying on forward sampling—has yet to emerge, primarily due to the monolithic, black-box nature of the environment model learned by neural networks \cite{mcmahon2022survey, otte2015survey}. Meanwhile, research in the planning literature itself that seeks to learn environment models often operates under highly restrictive assumptions \cite{jimenez2012review, mordoch2023learning, verma2021asking, stern2017efficient} that cannot hope to capture the expressivity of modern learning systems. Integrating the ability to learn robust environmental models with the capability to plan effectively and accurately based on these models would, hence, significantly enhance the capabilities of agential AI systems.

A class of methods that warrants mentioning in the integration of modern ML systems with classical approaches falls under the broad umbrella of \textit{neuro-symbolic AI}. These methods build upon research recognizing both the limitations of contemporary ML and the potential benefits of integrating them with human-designed algorithms demonstrating desirable properties like comprehensibility, speed and precision \cite{wan2024towards, colelough2025neuro}. However, the general approach in this field remains limited—artificial neural networks still serve as the fundamental building blocks of system design, treated as black boxes. As a result, the advantages of symbolic reasoning (e.g., comprehensibility, precision, and structured operational capabilities) do not extend to the internal representations learned by neural networks; rather, these benefits emerge only at higher levels of design, outside where neural networks operate within the system. This stands in contrast to a design paradigm that can maintain these desirable properties at every level of organization. As we will discuss in the remainder of this article, there is no intrinsic reason for this limitation—a proper AI system design paradigm, guided by principles that naturally result in a comprehensible and engineerable representation from the outset, can overcome it.

\subsection{Common origins}

What is the origin of these two major classes of problems in AI? Do they share a common foundation? It turns out that they, in fact, do; that being the lack of \textit{a structured representation} and the absence of \textit{the capability for learning in a structured manner}. To understand this, we must delve into the recent advancements in evolutionary theory.

\section{The Augmented Synthesis of Evolution}
\label{sec:evolution}

Evolutionary theory is usually regarded as only tangentially related to AI. However, upon closer examination, it becomes evident that it is arguably the most relevant of all, as this field has traversed a remarkably similar conceptual and intellectual trajectory to that which AI is currently navigating, albeit in previous decades.

\subsection{An overview of the Modern Synthesis}
\label{sec:modern_synthesis}

Evolution is the study of adaptive change in entities that possess the capacity for such transformations. The term is most popularly associated with evolutionary biology, which focuses on explaining how organisms have adapted—and continue to adapt—since the origin of life on Earth, enabling them to become increasingly suited to their particular environments. Modern evolutionary theory traces its roots to Charles Darwin’s On the Origin of Species \cite{darwin1964origin}, arguably the most significant intellectual work of the past millennium, influencing a vast array of disciplines from economics \cite{nelson2018modern} to physics \cite{zurek2009quantum} and engineering \cite{bartz2014evolutionary}. In this seminal work, Darwin proposes a common origin for all life on Earth, differentiating species through mechanisms of inheritable variation and adaptation to their respective environments via selection pressures. He correctly identifies the three main pillars required for evolutionary change: variation, inheritance, and selection \cite{marc2005plausibility, watson2016evolutionary}, and proposes a foundation for the latter through the mechanism of \textit{natural selection}, which continues to underpin our understanding of selective mechanisms in evolution to this day. Darwinian theories of selection, combined with our modern understanding of genetic variation and inheritance which traces its roots to Gregor Mendel (hence known as \textit{Mendelian genetics}) has formed the understanding of evolutionary biology that has dominated the 20\textsuperscript{th} century, which is known as the \textit{Modern Synthesis of evolution} \cite{huxley1942evolution, gilbert2000developmental, mayer2013evolution}.

In the view of Modern Synthesis (MS), inheritable genetic variation is essential for evolution and arises through mutations, which alter the DNA sequence, and through recombination, which reshuffles alleles during sexual reproduction. Without this \textit{genetic variation}, populations cannot respond adaptively to environmental changes. While many mutations are neutral or deleterious, some are advantageous and form the basis for beneficial \textit{phenotypic variation}. Selection is the differential reproduction of organisms based on their traits, where organisms with more advantageous traits in their respective environments can survive and reproduce more effectively. Selection, acting directly on the organismal phenotype—which is the visible expression of the organism's genotype—affects the survival and reproductive success of individuals. This process increases the frequency of advantageous alleles while reducing the prevalence of deleterious ones, facilitating adaptation to specific environmental pressures. Although selection occurs at the phenotypic level, the evolutionary changes manifest across the underlying genotype.

Population genetics is the study of how gene frequencies in populations change over time due to variation and selection mechanisms \cite{johnston2019population}. The Modern Synthesis uses population genetics as its backbone to explain evolution \cite{mayer2013evolution, gilbert2000developmental}, focusing on how allele frequencies change in response to natural selection and other evolutionary forces. It treats evolution as a process at the population level, abstracting away the complex internal processes of organisms and development, which aren't central to its explanation \cite{walsh2015organisms, gilbert2000developmental, marc2005plausibility, carroll2005endless, laland2015extended}. Biological evolution, in turn, is regarded as a property of populations, not organisms, and in the pure Modern Synthesis view, it is expected to manifest as slow, gradual changes in organisms across time \cite{carroll2005endless}.

\subsection{Limits of the Modern Synthesis}
\label{sec:ms_limits}

Modern Synthesis has been immensely successful in advancing our understanding of evolutionary forces from being nearly-absent in mid-19\textsuperscript{th} century to being a near-universally accepted theory by the end of 20\textsuperscript{th}. It has unified Darwin's theory of natural selection with Mendelian genetics, creating a comprehensive framework for understanding evolution. It established population genetics as the key discipline for studying evolutionary processes, clarified how natural selection leads to adaptation, and provided a model for speciation. Additionally, with its predictive power, it aided fields like medicine \cite{gould2009antibiotic, dolgova2018medicine}, agriculture \cite{thrall2011evolution}, and conservation \cite{olivieri2016evolution}. But as powerful as it is, the Modern Synthesis leaves gaps in its ability to explain aspects of evolution, whose importance, while downplayed in earlier 20\textsuperscript{th} century, are now understood to be much more important than initially thought.

First among these is one of the most crucial factors driving popular resistance to accepting evolution as a valid explanation for the origin of life: the fact that the Modern Synthesis does not incorporate (neither as explanandums nor as explanans) the generation of complex biological structures \cite{walsh2015organisms, gilbert2000developmental, marc2005plausibility, carroll2005endless} that clearly demonstrate identifiable features like repetition/reuse \cite{preston2011reduce,anderson2010neural, barthelemy1991levels}, correlation \cite{paaby2013many, watson2014evolution, price1992evolution}, modular organization \cite{lorenz2011emergence, clarke1992modularity, kadelka2023modularity, clune2013evolutionary, wagner2007road}, and hierarchical composition \cite{mengistu2016evolutionary, ingber2003tensegrity, uversky2021networks, grene1987hierarchies, pan2014exploring} in their phenotypes. Viewing evolution as a gradual statistical optimization of allele frequencies within a population, the Modern Synthesis explicitly excludes the internal complexity of organisms from its scope of inquiry. Evolution is framed as a statistical process over a population \cite{smith1978optimization}, with the organism treated more like a vessel for genes than a complex being with internal developmental processes \cite{walsh2015organisms}. However, it’s evident that for complete understanding of the evolutionary process, the aforementioned critical structural patterns in biological organisms (repetition, reuse, modularity, hierarchy) seen almost invariably across taxa — which ultimately influence their fitness and, in turn, loop back into the evolutionary process through selection — can’t simply be reduced to a set of continuous-valued parameters or binary variables. Yet, this is what a Modern Synthesis view suggests by asserting that the internal dynamics of an organism can be treated as a black box for the purpose of understanding evolution—implying that evolution adapts these structural properties as encoded in the symbolic string of genome. This is the gap between the genotypic variation and phenotypic variation \cite{marc2005plausibility}: While the Modern Synthesis satisfactorily explains the generation of genetic variation, the generation of phenotypic variation — the kind visible at the organismal level — is not explained MS. On the contrary, it is treated as a black-box process, supposedly not directly relevant to a full understanding of the evolutionary process.

The second limitation of the Modern Synthesis (intricately tied to the first, as will be clear shortly) is on the qualitative pattern of evolutionary progression that is predicted vs. observed. This is most popularly evident in the concept of punctuated equilibrium \cite{gould1977punctuated, marc2005plausibility}: Observations from the fossil record reveal that evolution doesn't proceed \textit{only} as slow, steady march as MS suggests. On top of these, we see periods of rapid change, where species undergo significant transformations in relatively short bursts, followed by long stretches of stability. An even more grandiose exemplification of this discrepancy between the observed pattern of progression and that predicted \& explained by Modern Synthesis, however, is highlighted by the exponential increase\footnote{Throughout the article, the term "exponential" is used conceptually for ever-accelerating processes driven by recursive improvement. As will be evident by the end of this section, this is the case for biological evolution as well.} in the speed of morphological change and the rise in phenotypic complexity \cite{smith1997major, koonin2007biological, russell1983exponential, sharov2006genome, heylighen2000evolutionary, kurzweil2006singularity} when we look at the evolutionary history as a whole, especially in broad cognitive capabilities of organisms. For instance, consider the timeline of life's complexity exemplified by the following landmarks \cite{newscientistTimelineEvolution, futuyma2013evolution} (see also Table 1 in \cite{gerhart2007theory}):

\begin{itemize}
    \item Emergence of life: Approximately 3.5-4 billion years ago, the first single-celled organisms appeared.
    \item First eukaryotes: Around 2-2.5 billion years ago, eukaryotic cells with more complex cellular structures emerged.
    \item First multicellular organisms: About 1 billion years ago, multicellular life forms began to evolve.
    \item First vertebrates: Approximately 500 million years ago, the vertebrates emerged.
    \item Emergence of Humans: \textit{Homo} genus appeared roughly 3 million years ago.
\end{itemize}

Despite these events being separated by comparable (even somewhat shortening) timescales, the phenotypic complexity and versatility of organisms have increased exponentially. The leap from single-celled to multicellular life, for example, involved a massive increase in biological complexity and evolution of associated specialization \& control procedures \cite{grosberg2007evolution,ispolatov2012division}, far surpassing the visible phenotypic change from prokaryotic to eukaryotic cells \cite{vosseberg2024emerging}; and likewise for the transition from first multicellular organisms to \textit{Homo sapiens} (or any other modern species with their wide repertoire of capabilities).

The difference between the progression pattern \& products of the evolution as observed qualitatively and what is predicted by Modern Synthesis is not only supported by retrospective analysis of the evolutionary history, but also experimentally in the field of computational optimization. Genetic algorithms \cite{sivanandam2008genetic} are a family of population-based optimization methods that mimic evolution based largely on a Modern Synthesis interpretation, as a straightforward optimization process within a predefined parameter space. While these algorithms are proficient at optimizing predefined structures (e.g. choosing the best combination of variables from a predefined set, or optimizing the continuous values for a set of design parameters provided \cite{bayley2008design, chiroma2017neural}), they are unable to spontaneously generate new structures within their design space, nor can they implement rapid, large-scale, and adaptive changes to their products. The only exception to this are a small class of methods called indirect encodings \cite{meli2021study, gauci2010indirect}, which mimic aspects of developmental processes and can sometimes produce interesting structures.\footnote{As visible from their nomenclature, even these methods exhibit the major misunderstanding of genome as an enconding of the structure of the organism, a point which we elaborate further in the following subsection.} Even then, the outcomes they generate are nowhere near the complexity and creativity of natural evolution, failing to approach the complexity or versatility of biological organisms produced by evolutionary process.

The core issue behind these shortcomings is that the Modern Synthesis explains \textit{why evolution occurs} and \textit{its general direction}, but not how biological organisms are produced. While MS offers a framework for understanding the conditions leading to phenotypic changes optimized for the environment, it mainly addresses the "how" questions through genetic variation—an essential prerequisite for evolution—by detailing mechanisms like mutation and recombination. However, genetic variation is just the starting point for phenotypic variation, which alone is subject to selective pressure.

What remains largely unexplained is the \textit{generation of phenotypic variation} and the \textit{mechanisms} behind the emergence of well-adapted structures in organisms. These structures often outperform the best human-designed machines, yet their production remains a mystery within the MS framework.\footnote{This limitation can also be framed through the three foundational pillars of evolutionary theory: variation, selection, and inheritance. The MS thoroughly addresses the latter two but provides a limited account of variation \cite{marc2005plausibility}. While it clarifies how genetic changes initiate variation, it treats the subsequent steps—how these changes lead to diverse phenotypes—as a black box, leaving much of the "how" of evolution unaddressed \cite{marc2005plausibility, laland2015extended}.} As will be evident in the next section, this gap in explaining how organismal structures emerge from genetic information is crucial for understanding the discrepancies between observed patterns of evolutionary progression and those expected by the Modern Synthesis.

\subsection{Evolutionary Developmental Biology}
\label{sec:edb}

Over the past few decades, advancements in various fields of biology have enhanced the evolutionary perspective provided by the Modern Synthesis, addressing aspects of evolution that were previously unexplained by or appeared to contradict the notion of statistical optimization across populations \cite{laland2015extended}. Among these advancements, Evolutionary Developmental Biology (EDB) has been particularly influential \cite{carroll2005endless, marc2005plausibility, west2003developmental, gerhart2007theory}. EDB focuses on the developmental mechanisms that translate a genome into a fully grown organism and examines how these mechanisms influence the evolutionary process—both in immediate, local contexts and in shaping the overall patterns of evolutionary progression.

Recent findings in Evolutionary Developmental Biology indicate that the gap between genotype and phenotype cannot be dismissed as a black-box process with irrelevant internal dynamics. On the contrary, the logic of development in biological systems exhibits universally shared properties that not only clarify the complexity and adaptability of organisms but also reveal significant aspects of the evolutionary process as a whole.

\subsubsection{The structure of gene regulation}
\label{sec:edb_structure_genereg}

The central insight of EDB is that the generation of phenotypic variation—the physical and structural traits of an organism that interact with the environment and serve as the direct targets of selection—is not solely driven by the genes and proteins an organism possesses. Rather, it is primarily shaped by changes in their regulation: specifically, by the spatial, temporal, and quantitative patterns of gene expression (i.e. “where,” “when,” and “how much” a gene is expressed) \cite{carroll2005endless}. This regulation is mediated by the genome's noncoding regions—segments of DNA \cite{shanmugam2017non} that do not code for proteins but play a crucial role in controlling the expression of coding regions. Historically, a large portion of these noncoding regions was thought to be nonfunctional "junk" DNA, but their significance in regulating coding DNA has become increasingly clear in recent decades \cite{makalowski2003not, carroll2005endless}.

Moreover, regulatory genes—such as transcription factors, microRNAs, and other molecular regulators—can control the expression of other regulatory genes, as well as the expression of coding genes (those that directly code proteins). This enables the hierarchical regulation of entire pathways of gene expression, where master regulators or key transcription factors initiate downstream cascades \cite{hansen2015effects}. This intricate network of interactions is not strictly hierarchical, as many regulatory pathways also feature feedback loops and non-linear connections \cite{barabasi2004network, holehouse2020stochastic, krishna2006structure, manicka2023nonlinearity}. This web of interactions, where different parts of the genome activate or suppress one another and coordinate downstream processes, is known as a \textit{gene regulatory network (GRN)} \cite{davidson2010regulatory,
levine2005gene}.

Small changes in regulation, when unfolded across time during development, can lead to dramatic alterations in the adult organism’s phenotype. A striking example of this is a famous experiment in \textit{Drosophila melanogaster}, where a small regulatory change resulted in the development of eye-like structures on the wings of the flies \cite{halder1995induction}. This occurred due to a mutation that misregulated the expression of a gene called \textit{Eyeless} involved in eye development. Normally, \textit{Eyeless} is expressed in the fly’s head to form the eyes, but in this experiment, a slight alteration in the regulation of this gene caused it to be expressed in the wing, triggering the formation of structures that resembled eyes—complete with photoreceptor cells.

This phenomenon highlights an important point in developmental biology: there is no need to evolve a new eye from scratch. Instead, by modifying the regulatory networks that control the expression of genes like \textit{Eyeless}, an entirely new structure (in this case, an eye-like structure) can be induced in a non-traditional location, such as the wing. The induced structure was surprisingly complete, exhibiting many of the characteristics of a true eye, such as photoreceptor cells and the typical eye morphology, albeit being nonfunctional, lacking the necessary connections to the nervous system and other supporting structures needed for vision. This serves to emphasize that the regulatory changes could re-deploy an existing developmental program to form a structure that could be integrated within the new location. The modification didn’t involve creating new genes but rather reshaping the expression patterns of existing genes, demonstrating the powerful role of regulatory networks in shaping developmental outcomes, corresponding to \textit{large qualitative changes in the phenotype} with \textit{very little modification at the genetic level} \cite{carroll2005endless}.

As highlighted earlier, biological organisms consistently showcase fundamental structural properties like modularity, hierarchy, and repetition \cite{preston2011reduce, barthelemy1991levels, lorenz2011emergence, kadelka2023modularity, clune2013evolutionary, ingber2003tensegrity, uversky2021networks, pan2014exploring}. They also reveal a striking correlation in changes across components \cite{paaby2013many, watson2014evolution, price1992evolution}. These traits stem from the architecture of the regulatory networks previously discussed \cite{carroll2005endless, alcala2021modularity, verd2019modularity, wagner2011pleiotropic}. Repetition arises from genetic networks that reuse similar regulatory mechanisms across various developmental processes. Modularity in an organism’s phenotype emerges when distinct gene sets regulate different parts of the organism, functioning largely independently. Meanwhile, hierarchical organization springs from higher-level regulatory genes controlling the expression of downstream genes, which govern sub-networks linked to specific modules or reused elements. Additionally, some components of gene regulatory networks are expressed in various regions of the organism, enabling changes in a single part of the genome to manifest in multiple areas of the phenotype, wherever the associated subnetwork is active. In sum, these structural properties of biological organisms reflect the nature and organization of gene expression within the genome (although the relationship is more complex than a simple one-to-one correspondence, as elaborated in the next subsection).

In addition to arising from the unfolding structure of GRNs during embryogenesis, these structural properties significantly influence how evolution alters organisms, facilitating the development of new functional structures while reducing the likelihood of detrimental changes:

\begin{itemize}
    \item Modularity and hierarchical organization facilitate both correlated changes—where modifications to one part of the organism trigger rapid, synchronous changes in others (which can be detrimental if the changes interfere with one another)—and isolated changes, allowing alterations in one component without affecting others \cite{carroll2005endless, verd2019modularity, clune2013evolutionary, mengistu2016evolutionary}. For example, when ancestral arms evolve into wings in flying animals like bats or birds, both wings develop in tandem due to shared genetic pathways. Conversely, isolated changes enable independent modifications in different body parts; for instance, transforming upper appendages into wings while preserving the functionality of the legs demonstrates how modularity allows for alterations in one region without disrupting others \cite{seki2012evolutionary, eliason2023early}.
    \item Repetition of substructures allows for the rapid generation of new structures from a moderately developed origin \cite{wagner2011pleiotropic}. A striking yet nonfunctional example of this principle is the formation of eye-like structures on the wings of \textit{Drosophila} discussed above \cite{halder1995induction}. A functional illustration would be the duplication of Hox gene clusters, which enabled the evolution of anatomical complexity and diversification in early vertebrates \cite{carroll1995homeotic}.
\end{itemize}

Another crucial feature of this system is the ability to preserve ancestral traits \cite{szilagyi2020phenotypes, carroll2005endless}. This is achieved by deactivating the switch points responsible for expressing these traits, which blocks their associated downstream genetic networks. This preservation does not completely dismantle the underlying genetic machinery, allowing for a simple reactivation of the trait later \cite{szilagyi2020phenotypes}. Thus, a seemingly small change in regulatory networks can rapidly lead to the re-emergence of a past, functional \& well-adapted trait, without having to entirely recreate the complex system that governs its expression.

\subsubsection{Process-encoding genome and local variation-selection}
\label{sec:edb_process_varsel}

The organization of gene expression as a regulatory network, featuring selective expression across various regions of the organism, explains the shared structural properties of modularity, hierarchical organization, and repetition seen across taxa. It also accounts for evolution's ability to enact both isolated and correlated changes within the organism's phenotype. However, this perspective only partially captures the complexities of development, as it overlooks a crucial aspect of organismal response: the adaptability and versatility of the generated structures.

Some structures or behavioral responses of an organism can be suitably generated and used as they are predictably throughout the lifetime of an individual in a species. It is plausible to think such structural properties to have been first generated \textit{de novo}, then optimized for adaptive perfection by the species-level evolutionary process — it is beneficial for the organism if such structures are well-evolved, fine-tuned and reliably generated. 

However, other structures or responses do not fit this category; they lack predictable, pre-determined shapes or patterns and must be generated in response to the demands of either the environment or other components of the organism. This generation can be specific to the individual organism—consistent throughout its lifetime—or in reaction to a changing and unpredictable environment encountered during its life. For the former, consider the development of muscles and veins, along with the subsequent innervation of an organism. Each individual within a species exhibits variable size—shaped not only by genetic factors but also by external conditions and resource availability \cite{brett1979environmental, lynch2012effect}—which necessitates unique development of these support structures tailored to the bone size and shape of that specific organism. The latter, on the other hand, is exemplified by the organism's immune responses, which must adapt to the specific pathogens it may encounter throughout its life. Similarly, animal behavior is influenced by immediate environmental needs, requiring responses that are directly modified by changing circumstances.

The important point in understanding how these types of adaptively generated structures can be formed, evolved, and in turn affect the further evolution of the organism lies in the understanding that the genome does not act as a blueprint or even as an encoding of the organism, as was thought earlier in the 20\textsuperscript{th} century \cite{natureItsTime}. Instead, it encodes the \textit{cellular processes} that, throughout unfolding, modify the structure, responses, and behaviors of the cells that express them—less like a blueprint and more like a control system or an algorithm for the lowest-level agent, the cell. As such, the generation of the organism's structure is less a process of decoding an encoding represented in the genome and more a not-necessarily-predetermined outcome of the interactions of these processes, encoded in the genome, with each other and (importantly) with the environment \cite{marc2005plausibility, gerhart2007theory, west2003developmental}.

The significance of this scheme of structure generation through responsive processes—especially in achieving reliable development amidst changing circumstances and hence facilitating evolutionary change—is clearly demonstrated by the development of connective tissues associated with the skeletal system, particularly the coordinated development of muscles, blood vessels, and nerves around bones. During tetrapod limb development, the initial limb bud contains only precursors for bone and dermis. The muscular, neural, and vascular systems then adaptively develop around this foundation \cite{gerhart2007theory, marc2005plausibility}:

\begin{itemize}
    \item Muscle precursors receive signals from the developing dermis and bone, guiding their positioning relative to these structures \cite{kardon2003tcf4, christ2002limb}.
    \item Axons extend in large numbers from the nerve cord into the limb bud. Some make contact with muscle targets and stabilize, while the rest retract \cite{li2012regulation}. This process of initial overproduction followed by selective stabilization is a recurring feature of neural development throughout the body \cite{hiesinger2021self}, as discussed further below.
    \item Cells in the developing limb secrete signals in response to oxygen demands, inducing nearby blood vessels to extend toward them \cite{adair2010angiogenesis, ferrara2003biology}. This mechanism also mirrors similar processes that drive vasculogenesis all throughout the body \cite{gerhart2007theory}.
\end{itemize}

The importance of this is that an evolutionary (or in some cases, a somatic) change in the bone structure of an organism, such as an elongation of the dimensions that is beneficial to the environment or even the formation of a new appendage, does not require a coordinated change in the associated muscular, vascular and neural structures in a hypothetical blueprint of the organism at the genetic level \cite{marc2005plausibility, gerhart2007theory}. The same would be true even in larger-scale changes, such as the formation of additional appendages altogether \cite{malik2014polydactyly}.  These associated changes are readily accounted for in the developed organism, since these structures are generated in response to the skeletal system’s structure in the first place \cite{west2003developmental}. As such, an adaptive change can take place rapidly and safely with the relevant mutations that affect the skeletal system; without requiring any change in how these associated tissues are generated.

Adaptive growth of morphological structures like skeletal or muscular system is an important functionality for all organisms. Yet these are still structures that, under normal circumstances, are created and used throughout the life of an organism, and are not particularly responsive to rapid changes required by the environment. An even more striking example of adaptive somatic response within organisms is found in the immune system, which requires a type of reactivity that operates on a timescale much faster than the organism's lifetime. The clonal selection theory of the immune system \cite{burnet1957modification, burnet1957modification}, still a foundational concept in immunology \cite{rajewsky1996clonal}, proposes an adaptive response for both the immediate reaction of the immune system and the establishment of long-term immunological memory. According to this theory, when an antigen—foreign molecules or structures, such as proteins on viruses or bacteria that trigger an immune response—enters the body and binds to a lymphocyte (a type of white blood cell) with surface receptors that match the antigen, the lymphocyte activates and proliferates into a diverse array of descendants. These descendants vary from the parent cells due to mutations, refining their ability to bind to the antigen. This selective proliferation, based on affinity to the antigen, ensures that the immune system evolves in real time to efficiently combat invaders while retaining memory for future immunity. In this way, the immune system can respond effectively to foreign agents in the organism’s immediate environment, even though the specific antigens requiring a response are unknown in advance and cannot be preprogrammed by evolution. Its capability to recognize and react to a vast range of antigens through random variation and refinement on a somatic level enables this adaptability.

The operation of the immune system exemplifies a general class of mechanisms often referred to as \textit{exploratory processes} \cite{gerhart2007theory, marc2005plausibility}, which are employed to generate highly adaptive responses in organisms. These processes leverage the variation and selection principles underlying biological evolution at the population level in a localized manner, where variants of the same substructures are generated somatically and then refined or retained through a selective signal. The clonal selection theory discussed earlier is an example of such a process, where variation occurs among lymphocytes and the selective signal is their affinity to the antigen. This affinity is crucial for further proliferation; lymphocytes that do not meet this criterion are naturally discarded over time.

Similar principles are evident in the development of the neural system, including the brain \cite{hiesinger2021self, marc2005plausibility}. During development, neurons initially form more synaptic connections than necessary (synaptic overgeneration). This excess undergoes a selection and pruning process, where the strongest and most effective connections are retained while weaker ones are eliminated \cite{bourgeois1989synaptogenesis, rakic1986concurrent, chechik1998synaptic, sakai2020synaptic, petanjek2023dendritic}. Here, variation is based on synapses, and selection occurs over connection strength. A similar mechanism operates during axonal growth: neurons extend axons to target cells, initially forming many branches. Like synaptic overgeneration, excess branches are pruned through a selection process that strengthens effective connections and retracts or eliminates ineffective ones \cite{lamantia1990axon, rakic1983overproduction, provis1985human, kaiser2009simple, innocenti1997exuberant}.

These processes, based on somatic and local variation and selection, are crucial for adaptive responses in scenarios where there is little prior knowledge of the responses that will be needed by the organism in its lifetime, as illustrated by the two examples given. This principle is a universal mechanism for the realization of adaptive responses and versatile structures in such scenarios, showing that the Darwinian variation and selection mechanism for the generation of adaptive structures operates not only at the population level or on the basis of the genome, but also as a unifying force across all biological systems in critical scenarios where responses cannot be predetermined in advance (which naturally includes population-level evolution as well) \cite{marc2005plausibility}. It is also noteworthy that the process of local variation and selection, through its influence on the neural system, also underpins intelligence itself — a point that will be important and further elaborated in Section \ref{sec:is_intelligence_evolutionary}.

\subsubsection{Implications of EDB for understanding the progression of evolutionary process}
\label{sec:edb_understanding}

The implications of EDB for our understanding of evolutionary progression and the evolutionary process suggest that large portions of the expressed genome—specifically coding DNA, which encodes proteins, as well as the associated low-level regulatory networks—are highly conserved across distantly related organisms \cite{marc2005plausibility, gerhart2007theory}. This includes the fundamental processes responsible for generating basic cellular structures, as well as higher-order structures and exploratory processes that allow organisms to generate adaptive responses based on local variation and selection. Such conserved \textit{core processes} (as termed in \cite{marc2005plausibility} and we adopt) evolved early in evolutionary history and have been preserved across vast phylogenetic distances. Examples of these conserved processes include the genes which govern body plan development in all animals \cite{lemons2006genomic}, the signalling pathway which plays a crucial role in cell differentiation across metazoans \cite{holstein2012evolution}, and the cytoskeletal components which form the structural backbone of cells across eukaryotic life \cite{erickson2007evolution}. The fundamental mechanisms of gene transcription \cite{werner2011evolution} and core metabolic pathways \cite{fothergill1993evolution} are also highly conserved, indicating their early emergence and essential role in life’s functioning.

By contrast, major evolutionary changes since these early developments have primarily occurred at the level of regulatory networks, which control the expression and interaction of this foundational genetic toolkit. Examples of such high-level regulatory changes include differences in regulatory elements governing limb development in vertebrates \cite{burgess2016sonic, tickle2017sonic}, the rewiring of neural gene expression networks that led to expanded brain function in humans \cite{wang2016divergence, patoori2022young, suresh2023comparative}, and the evolution enhancers that influence gene expression patterns for morphological novelties in different species \cite{rebeiz2017enhancer}. These regulatory modifications not only drive organismal diversity but can themselves become conserved and serve as the substrate for further evolutionary change over time\footnote{Nonfunctional examples of major organismal changes driven by regulatory alterations can also be provided, such as the human birth defects resulting in extra fingers \cite{lettice2008point}.)}.

Some scientists take this even further, interpreting phenotypic variation as being generated in a biased, non-random fashion \cite{marc2005plausibility, gerhart2007theory}. In their view, while genetic variation primarily arises through the random processes of mutation and recombination, phenotypic variation—i.e., the traits expressed in the adult organism—is shaped by past evolutionary successes. This variation tends to favor the reuse of conserved processes that have proven beneficial, it is also biased towards changes that can be easily integrated into the existing biological framework, taking advantage of the adaptiveness of peripheral processes and structures. Furthermore, the generation of phenotypic traits is not simply predetermined but actively shaped through exploratory processes, where the organism generates structures adaptively in response to its specific environmental and physiological needs, rather than in a strictly deterministic fashion.

Recall how, as we discussed earlier, the statistical-optimization model of the Modern Synthesis struggles to fully account for the complexities of evolutionary history. This view, while useful, falls short when it comes to explaining phenomena like punctuated nature of evolutionary history (periods of stasis interrupted by rapid bursts of diversification and complexification) and the exponential increase -accelerating with recursive improvement- in phenotypic complexity and organismal capability. A view of evolution that incorporates the influence of developmental processes and their inherent properties, on the other hand, offers a satisfactory explanation. It accounts for bursts of complexification and rapid evolutionary changes, building on the latent potential shaped by conserved processes honed through evolutionary history.

At the heart of this is the concept of \textit{conserved core processes} we discussed above: Once core processes are fully evolved and optimized, they can be reused (as processes) or repeated (as structures) when needed, with minimal genetic changes focused mainly on the regulation of these processes \cite{marc2005plausibility, gerhart2007theory}. This ability to re-deploy core processes is facilitated by the adaptability of peripheral systems—take the previous example of the adaptive development of muscular, vascular, and neural systems surrounding skeletal structures—which require minimal modifications, despite large phenotypic differences. Furthermore, a set of pre-existing, stabilized core processes that underpin fundamental functions can be combined, rearranged, or reconfigured to produce more complex functions or structures at higher levels, as exemplified in the evolution of the vertebrate eye \cite{lamb2007evolution, lamb2008origin}, which involved the repurposing pre-existing components like the photoreceptor cells (originally served simple light detection for purposes like phototaxis), lens (originally derived from the dermis) and the visual neural circuitry (derived from the ancestral sensory neural circuitry). The evolution of such structures requires far fewer changes compared to evolving them from scratch, as the organism is already working with well-established processes that can seamlessly integrate with new adaptations. Alongside these, processes that don’t culminate in final structures but instead drive exploratory processes—such as immune system responses \cite{rajewsky1996clonal, burnet1957modification, burnet1957modification} and neurons forming extensive connections \cite{hiesinger2021self, bourgeois1989synaptogenesis, rakic1986concurrent, chechik1998synaptic, sakai2020synaptic, petanjek2023dendritic, lamantia1990axon, rakic1983overproduction, provis1985human, kaiser2009simple, innocenti1997exuberant}, as previously discussed—allow the organism to co-opt these mechanisms for adaptation, not just across generations, but within its own lifetime. This enables real-time responses to environmental pressures without the need for prior structures tailored to the specifics of their circumstances, allowing organisms to exponentially increase their capabilities once a solid foundation (e.g. a neural system capable of reconfiguration and learning) is in place, without relying on explicit evolutionary changes (e.g. the evolution of complex pre-coded behavior patterns) to fully harness these newfound abilities. This view explains why periods of rapid increases in complexity and capability are a persistent phenomenon throughout evolutionary history, and why these increases seem to accelerate exponentially when viewed across the entire evolutionary timeline.

This description of the evolutionary process not only offers a more comprehensive view than one based solely on the Modern Synthesis—which primarily emphasizes population-level statistical optimization—but also renders the entire process more intuitive. It reflects the mechanisms and qualitative properties of other progressive processes we are intimately familiar with, at least regarding their outcomes, such as learning in humans and technological advancement. In these processes, the conservation of early gains, their preservation, and their utilization to drive further progress at an ever-increasing pace are both inherent and natural. This parallel between EDB and these domains will be central to our discussion in the following sections.\footnote{In Evolutionary Developmental Biology, there is also a very central are of inquiry which focuses on how developmental mechanisms and phenotypic responses to environmental influences can loop back into inheritable traits, or facilitate further evolution at the genomic level. While we won’t delve into this aspect, as it’s not directly relevant to the argument in this paper, interested readers are referred to works on developmental plasticity \cite{west2003developmental} and epigenetics \cite{allis2015epigenetics}.}

\subsection{Modern Synthesis and contemporary AI}
\label{sec:ms_ai}

The Modern Synthesis of evolution and the current state of machine learning (ML) share striking parallels. Both view their respective processes of adaptation—evolution and learning—as fundamentally statistical optimization processes. They emphasize a numerically defined, averaged final performance (fitness in evolution vs. accuracy/reward/error in ML) rather than focusing on the immediate responses of agents. Both approaches treat their subjects as largely unstructured, encapsulating relevant properties in abstract continuous variables—gene frequencies in biology and edge weights in ML—without addressing the deeper meaning behind these variables. While each has been successful in its own right—Modern Synthesis offering a substantial explanation of evolution and ML solving complex high-dimensional tasks—both are considered incomplete for similar reasons: (1) a lack of "structure", (2) the pattern of progression of evolution, as explained by the Modern Synthesis, and of learning, as realized by ML.  Modern Synthesis, while not inherently contradictory to extensions that address it, does not fully explain certain aspects of evolutionary progress, namely the speed of adaptation, the inherent adaptability of organisms, or the widely-observed structural properties of biological systems. Analogously, contemporary machine learning faces criticism for its data inefficiency ($\sim$speed of adaptation), inability to leverage past knowledge when learning new tasks without sacrificing previously acquired information ($\sim$inherent adaptability), and opaque internal structure ($\sim$structural properties).

The resemblance between the previously dominant view of evolution and contemporary AI, particularly regarding their limitations, raises an intriguing question: Can the principles behind the recent shift in evolutionary theory—largely driven by Evolutionary Developmental Biology, whose core ideas we explored in this section—also form the foundation for improving AI? At this point, this resemblance primarily seems as a compelling analogy. Before we explore whether this expanded view of evolution could catalyze the next paradigm shift in AI, we will first engage in a deeper analysis that reveals a deeper relationship between intelligence and evolution, one that transcends a simple analogy.

\section{Is Evolution Intelligence?}
\label{sec:evolution_intelligence}

The analogy and resemblance between intelligence and evolution extend beyond the conceptual stages, assumptions, and capabilities of their respective "theories"—the Modern Synthesis of evolution and the contemporary machine learning paradigm based on function approximation with large unstructured networks, as discussed in Sections \ref{sec:contemporary_ai} and \ref{sec:evolution}. This resemblance runs deeper, suggesting even the potential for equivalence between the two processes at a higher level. In this section, we delve into the more speculative aspects of this relationship from two directions: First, we examine a recent information-theoretic analysis of the phenomenon of life and how a crucial defining property of life, and possibly of evolution, corresponds to a plausible definition of intelligence. Next, we discuss how our current understanding implies the presence of evolutionary mechanisms within the biological structures that underpin intelligence, as well as the necessity of these evolutionary mechanisms for learning in an environment characterized by uncertainty about the future.

\subsection{Is evolution intelligent?}
\label{sec:is_evolution_intelligent}

A defining characteristic of living systems is their ability to locally resist entropy increase—a fundamental measure of disorder—by preserving internal order. They accomplish this by functioning as open systems, continuously exchanging energy and matter with their environment. This dynamic allows them to sustain low internal entropy while increasing the entropy of their surroundings. This near-universal trait is often considered a plausible definition of "life" \cite{schrodinger1946life, von2019ab}. A recent line of research \cite{parr2022active, friston2013life, friston2019free, sella2005application} analyzing life through the lens of information theory and Bayesian inference reveals a striking connection between this fundamental property and what can be described as intelligent, anticipatory behavior \cite{rosen2011anticipatory}. Here, we provide an overview of the central argument of this \textit{free energy principle} as it pertains to our inquiries, directing the reader to \cite{parr2022active, friston2013life, friston2019free} for a full mathematical treatment.

Starting from the premise that organisms must maintain internal order by minimizing entropy increase\footnote{Formally, this principle is captured by modeling organisms as entities enclosed by Markov blankets \cite{friston2013life}. The persistence of this boundary necessarily entails minimizing internal entropy \cite{friston2019free}, expressed as the Shannon entropy of the organism’s generative model, which serves to explain its observations. Throughout this section, entropy refers to Shannon entropy,  $H(X) = -\sum_i p_i ln(p_i)$ (where $p_i$  is the probability of a given value $i$ for random variable $X$). This measure corresponds to thermodynamic (Boltzmann) entropy, $S = k_b ln(\Omega)$ when the system consists of $\Omega$ microstates with equal probability (see also \cite{karnani2009physical}).}, it is shown in \cite{parr2022active} that this process can be expressed as the minimization of a quantity called \textit{free energy} of the generative model of the organism— a model within the mathematical representation of the organism, resulting from the organism's separation from its environment by a Markov blanket \cite{parr2022active, friston2013life}, explaining and predicting the organism's observations and effectively representing its environment. Free energy, in turn, serves as an upper bound on entropy of this model.

It is demonstrated in \cite{parr2022active} that minimizing free energy—given an internal model of the environment and evaluated over extended timescales (i.e., in expectation)—can be broken down into two fundamental drives of the organism: (1) ensuring consistency between its internal model and the observed data from the environment, and (2) seeking observations that maximize information gain for future adaptation. Moreover, the first drive—achieving model-observation consistency—can be realized in two ways: either the organism updates its internal model to better predict incoming observations or it preferentially seeks observations that align with its existing model through its actions. The latter approach also involves a preference for internal states that are beneficial to the organism, encoded as \textit{priors} within the generative model.

Thus, the core activity of internal entropy minimization—essential for the survival of any organism (and life in general)—involves the combined effect of three crucial processes:

\begin{itemize}
    \item Modifying the model to align with observations (perception and learning),
    \item Preferring observations that closely match the model (pragmatic action),
    \item Seeking observations that maximize adaptive changes to the model (information gain).
\end{itemize}

Although the definition of intelligence is not universally agreed upon \cite{legg2007collection}, and while this definition can always be refined and nuanced, we hold the view that the three points outlined above can serve as a plausible foundational definition of intelligence, one that is neither overly simplistic nor limited to human-level cognition.\footnote{We explicitly note that we are not considering consciousness and its role in intelligence in this inquiry, and when we refer to intelligence, we are considering a potentially non-conscious process. It is our view that consciousness is not a well-understood phenomenon neither in terms of its physical and biological underpinnings nor its functional role in (advanced) intelligence \cite{chalmers2017hard, ginsburg2019evolution}; hence, its discussion falls well outside the scope of this work.} An agent capable of learning a model and following the principles encoded within it—principles that encompass both its genetically determined preferences and its drive to engage with environments or states that closely match the model (i.e., environments where the agent is well-adapted)—can be regarded as displaying intelligent behavior. Furthermore, if the agent is also capable of recognizing the value of new observations for enhancing its model, performing guided exploration even when such actions do not immediately benefit its survival, it can be considered to be exhibiting intelligent behavior under all but the most restrictive definitions of intelligence (which often limit intelligence to humans). This formulation thus suggests that intelligence is an inherent consequence of the fundamental activity of internal entropy minimization in life. While this may blur the line between intelligence and life itself, such a merging can be useful. It leads us to a deeper understanding of the shared principles underlying both, which is particularly valuable given that we have a clearer understanding of life than intelligence, at least in terms of their fundamental natural principles. 

How, then, does this connect to the broader scope of evolution? Can this analysis—drawing an equivalence between a fundamental activity of life and a plausible definition of intelligence—be extended to encompass the entire biosphere, with evolution as the governing process? 

An argument can certainly be made: In a general sense, biological evolution -as the unfolding of the general principles of variation, selection, and inheritance on biological systems on a population level- can be defined as the process that governs the dynamics of the living part of the biosphere on Earth (influenced by, but not directly governing, the non-living factors of the ecosystem as well). The living part of the biosphere on Earth is formed of all the organisms that we can define as “alive.” As we know that each individual living being acts to reduce or keep low its own internal entropy, we can say that the alive part of the biosphere also reduces its entropy as a whole.\footnote{Formally because entropy is additive in independent systems \cite{shannon1948mathematical}, and multiple organisms (e.g. all living beings on Earth) have their internal states conditionally independent given their environment (e.g. non-living part of the biosphere) since the organisms can only interact via their environment (based on the definition of an organism over the presence of a Markov Blanket \cite{friston2013life, parr2022active}).} This, together with our initial definition of biological evolution as the process that governs the dynamics of the system formed by all biological organisms on Earth, corresponds to the corollary that biological evolution is a process that results in the reduction of the entropy (or countering of the increase of the entropy) of the open system of its operation (composed of all living beings on Earth).\footnote{This specifically relates to the information-theoretical entropy, defined in terms of the model evidence of the system composed of the living part of the biosphere \cite{parr2022active} and their collective observations. This often results in the complexification of the system model, which predicts these observations—the model corresponding to the totality of biological life—to better represent the environment. Such complexification can increase the internal entropy of the system's components (e.g., genetic diversification) while simultaneously reducing the surprise (and thus information-theoretical entropy) of the overall system regarding its observations from the external environment, akin to how an advanced neural system like those in primates can have higher internal complexity compared to simpler taxa, yet in doing so provide a more accurate representation of the environment, resulting in lower Shannon entropy. In other words, the entropy minimization argument holds valid only for the Shannon entropy of the observations of life as a whole from its external environment, distinguishing it from other definitions of entropy that may describe the internal composition of biological life, which need not be minimized. The former (Shannon) is the relevant type of entropy in our consideration of a potential equivalence with intelligence in this section.} The argument above regarding the equivalence of an activity of minimization of internal entropy and a plausible definition of intelligence would, in this interpretation, apply to the alive part of the biosphere and, as a corollary (as the highest-level process that governs the dynamics of this collective system, invariably across species and since the emergence of life, to the best of our current knowledge), to the process of evolution as well.

The idea that evolution can be reasonably characterized as an 'intelligent' process is not unique to us. Beyond the argument we outlined above, the original developers of the free energy principle have explored extending their framework to encompass evolution as a whole \cite{friston2013life}. Further work has explicitly framed natural selection itself in terms of free energy minimization \cite{sella2005application}. In addition to the free-energy perspective discussed throughout this section, a different line of research has also drawn conceptual parallels between learning theory and evolutionary dynamics \cite{watson2016can, power2015can} and established a formal equivalence between replicator dynamics in evolutionary theory and Bayesian inference (the foundation framework of the free energy principle) \cite{czegel2022bayes, czegel2019multilevel}.

Would this interpretation be valid, and can evolution be regarded as a process that can plausibly regarded as intelligent, based on a definition of intelligence that is neither too simplistic nor too anthropocentric? We refrain from making definitive statements on this question, as the topic, especially our discussions as they pertain to the totality of biosphere and the dynamics of their evolution are inherently speculative and necessitate a more in-depth, formal analysis that exceeds the scope of this work; for now we leave it to the reader's judgment and interpretation. However, we acknowledge it as a plausible perspective that could offer insight into the deeper relationship between evolution and intelligence, tracing a path from the fundamental principles of life to the a reasonable definition of intelligence. To explore the relationship in the opposite direction, we now turn our focus to the known processes of intelligence as a starting point.

\subsection{Is intelligence evolutionary?}
\label{sec:is_intelligence_evolutionary}

In Section \ref{sec:edb_process_varsel}, we discussed exploratory processes—mechanisms operating within the organism that rely on somatic and local variation followed by selection. These processes exemplify the widely-seen phenomenon of Darwinian evolutionary mechanisms at a sub-population scale \cite{gerhart2007theory, marc2005plausibility, west2003developmental}, enabling the organism to generate adaptive responses to unpredictable circumstances. Such responses cannot be pre-encoded efficiently within the organism through genetic evolution, given the longer timescales required for population-level evolutionary change.

As discussed above, exploratory processes are widely observed in the formation and adaptation of the nervous system. Synaptic and axonal overproduction, followed by selective retention, occurs extensively during development and childhood \cite{hiesinger2021self, nihDevelopingBrain, huttenlocher2013synaptogenesis}. These processes persist into adulthood under conditions demanding high plasticity, albeit on a much smaller scale and in more localized regions \cite{gu2013neurogenesis, gonccalves2016vivo, mowery2023adult}. The widespread occurrence of these Darwinian mechanisms in neural development suggests they play a critical role in the emergence of intelligence.

Yet, this is not the whole story of evolutionary principles and their role in brain function. Since the mid-20\textsuperscript{th} century, numerous theories have been proposed to explain not only the local adaptive behavior of neurons but also higher-order brain functions through evolutionary principles. Among the most well-known is the theory of neural group selection \cite{edelman-neuraldarwinism-1987, edelman1993neural}, which posits a two-stage variation-and-selection process: first across neuronal connections and then across neuronal groups. Another line of research theorizes that neuronal activity patterns can implement evolutionary replicator dynamics \cite{fernando2010neuronal, de2015neuronal, fedor2017cognitive}, while other theories propose that synapses themselves act as self-interested agents \cite{seung2003learning} or as units of selection akin to genes \cite{adams1998hebb}, alongside additional formulations that explore alternative selectionist frameworks \cite{changeux1973theory, loewenstein2010synaptic, calvin1987brain, calvin1998cerebral, fernando2012selectionist}. While these ideas remain relatively speculative due to both the difficulty of experimental validation and the broad, sometimes imprecise, nature of their formulations, they nonetheless stand as plausible candidate theories for explaining brain function. This plausibility is reinforced by the well-documented presence of local evolutionary mechanisms at the neuronal level (such as synaptic and axonal overproduction followed by pruning, as described above) and the selectionist characteristics observed in certain high-level cognitive functions—such as selective attention \cite{johnston1986selective}.

At this point, a reader familiar with modern machine learning methods might assume that mechanisms based on local variation and selection—clearly evident in the brain and potentially crucial for intelligent behavior—are absent from even the most advanced machine learning models. However, this is not the case. While we are not aware of any deliberate application of the principle of local variation and selection in the design of neural network-based machine learning systems, a direct connection between mainstream machine learning techniques and internal evolutionary processes can be drawn from at least two perspectives.

The first perspective involves reinforcement learning \cite{Sutton1998}, a widely used approach for learning from external environmental rewards, particularly in agent-based learning systems. Notably, \cite{borgers1997learning} demonstrated that in the continuous-time limit, the learning model converges to the replicator dynamics of evolutionary game theory \cite{smith82}. This establishes a formal equivalence between one of the fundamental frameworks of modern machine learning and a foundational principle of evolutionary theory, later augmented by the previously-mentioned works that linked the same principle to Bayesian inference as well \cite{czegel2022bayes, czegel2019multilevel}.

The second perspective emerges from overparameterized neural networks trained via gradient descent. These networks can be interpreted as undergoing a selection process over an initially abundant variation—represented by randomly initialized weights. Gradient descent functions as a selective force, amplifying "beneficial" weight patterns that reduce the error while diminishing less optimal ones, akin to natural selection refining traits in a population. Ultimately, this process converges to a local minimum, where the most "fit" weight configurations are selected and stabilized, making the training process conceptually analogous to evolutionary selection. Crucially, in this view, the variation-selection cycle is not iterative; there is no inherent mechanism for regenerating variation atop existing structures. Instead, the process is a strong, highly directed selection mechanism (minimization of a cost function) acting on an initially vast variation pool (the overparameterized, randomly initialized network). As training progresses, this pool is reduced to a fully stabilized network (a converged solution), where the remaining weight patterns are non-random and functional, actively contributing to the network’s learned function approximation task. Importantly, in this process, there are \textit{no means of introducing new variation} without disrupting at least some of the existing knowledge, as selective pressure—via gradient descent—affects the entire network uniformly.\footnote{Note that stochastic gradient descent is no exception to this, as it only introduces minor fluctuations in the direction of selection across training batches, rather than generating new variation among parameters.} This perspective not only offers a fresh interpretation of why neural networks possess such remarkable expressive power—an immense reservoir of variation honed by a potent selective signal—but also sheds a new light on the \textit{destructive adaptation} problem in continual learning (i.e., the inability of neural networks to acquire new knowledge without destroying existing structures) discussed in Section \ref{sec:ai_destructive_adaptation}. Fundamentally, the issue stems from the network’s inability to generate new variation locally when and where needed, without annihilating prior weight patterns. This stands in sharp contrast to exploratory processes in biological systems, which can locally generate variation on demand (see Section \ref{sec:edb_process_varsel} and preceding discussion), teasing one of the major points regarding the future of AI we will discuss in the following section.

To conclude, evolutionary mechanisms undeniably play a fundamental role in the emergence of intelligence in biological systems and are also implicitly embedded in the most effective artificial learning methods employed today. While we do not yet fully understand how these processes integrate into higher-level brain function or the extent to which they determine complex cognitive processes, we do know that somatic evolutionary mechanisms play an indispensable role in at least some aspects of cognition—particularly in learning. This necessity arises from the fundamental uncertainty regarding the future that any intelligent agent—whether an organism or an AI system—inevitably faces. For any given observation or circumstance encountered at a particular moment, the agent can generate multiple plausible solutions or internally consistent models/explanations, as its representational capacity inherently exceeds the requirements of any single observation (given that its cognitive model must accommodate more than just the present instance). Among these alternative solutions, many may be \textit{neutral} with respect to each other \cite{wilke2001adaptive, tenaillon2020impact} meaning they perform similarly well in the present context but differ in their suitability for future scenarios. Since an agent cannot a priori determine which of these neutral alternatives will prove most beneficial in the long run, the natural resolution to this problem is a mechanism of variation and selection. Multiple solutions of comparable immediate utility are generated and explored in parallel \cite{wagner2011origins}, and over time, those that prove most effective in future contexts are retained, while the rest are discarded.

\subsection{Summary}

In this section, we examined the intricate relationship between evolution and intelligence from two perspectives. First, we considered the connection from evolution to intelligence, exploring the possibility that a fundamental property of evolution can be decomposed into components that plausibly constitute a general definition of intelligence. Second, we analyzed the link from intelligence to evolution, highlighting the presence and importance of internal processes that embody fundamental evolutionary principles within both natural and artificial intelligent systems.

This deeper connection between evolution and intelligence reinforces our argument that beyond the conceptual analogies drawn between classical evolutionary theory and contemporary machine learning, there are even stronger reasons to believe that the conceptual advancements that addressed the limitations of the former can similarly be leveraged to overcome the limitations of the latter.

In the next section, we will discuss these principles in detail and examine how they can be translated into concrete design principles for the next paradigm in AI system development.

\section{The Future of AI}
\label{sec:ai_future}

\subsection{Principles for learning structured representations without destroying past knowledge}
\label{sec:ai_future_principles}

Previously, we discussed the connection between evolution and artificial intelligence from three different perspectives. The first of these perspectives, which is also the most direct and pragmatically relevant in suggesting a future path for AI, was the analogy between the dominant 20\textsuperscript{th}-century view of evolution (Modern Synthesis) and contemporary machine learning (deep, overparameterized neural networks optimized by gradient descent), considering both their successes and limitations. The remaining two—the consideration of evolution as a process that can potentially be decomposed into a plausible definition of intelligence and the examination of somatic evolutionary processes underlying the biological implementations of intelligence in nature—primarily served to deepen this analogy; yet at the same time, they signaled the possibility of an even more profound relationship between evolution and intelligence.

Armed with this connection, we now examine the principles behind recent advancements in evolutionary understanding beyond the Modern Synthesis, primarily driven by developments in Evolutionary Developmental Biology (EDB), and explore how they can address the parallel limitations of contemporary machine learning—most notably, its inability to acquire new knowledge without erasing existing knowledge (continual learning without destructive adaptation) and the unstructured, often incomprehensible nature of its internal representations. EDB has addressed the parallel limitations of the Modern Synthesis—specifically, its inability to explain the commonly observed structural properties of biological organisms and patterns of rapid adaptive changes in evolutionary history—by recognizing the complexity of the core control processes at the lowest levels of biological organization. This perspective incorporates structure through mechanisms such as encapsulation, modularity, hierarchy, and reuse, which emerge from the genome's structure as regulatory networks unfolding during development. Additionally, EDB identifies key principles of these low-level processes, notably local variation and selection at the somatic scale, which account for the adaptive generation and modification of structures or behaviors in scenarios where no perfect evolutionary prior is possible.

The same "design principles" can overcome the limitations of AI and lay the groundwork for a new paradigm of learning that generates comprehensible models in a structured representation and is not forced to damage or obliterate existing knowledge when learning something new.

We identify the following principles as being primarily important and integrable into AI (categorization being highly interdependent and mutually non-excluding):

\paragraph{Encapsulation and Core Processes:} To ensure stability in AI architectures, just as biological systems encapsulate core processes to facilitate reuse and prevent interference, learning should focus on the acquisition/generation and encapsulation of fundamental processes such as perceptual representations or behavioral patterns. These core processes, which remain conserved over time, serve as stable building blocks for more complex functions and are interconnected through a weak linkage \cite{marc2005plausibility, gerhart2007theory}, enabling the preservation of internal structures and recombination of learned functions.\footnote{Note that, beyond its relevance to developmental biology, this point also ties into the mechanisms underlying major evolutionary transitions \cite{smith1997major} and the emergence of new levels of selection—an aspect we have not explored in depth here but one that serves as a foundational pillar in recent advances in evolutionary theory.} Structural development in AI should emerge from encapsulation dynamics, where conserved, frequently reused processes form the foundation for increasing complexity. Evolutionary patterns prevalent in biological systems, such as the mechanisms of developmental modularity \cite{bolker2000modularity, wagner2007road} and duplication \& differentiation \cite{taylor2004duplication, copley2020evolution, minelli2000limbs}, can realize the targeted modification of parts of the previously-learned structures while conserving the rest, or the conservation of the original form's adaptive processes while generating their variants tailored for different purposes. Furthermore, this organically evolving structure with a modular and hierarchical composition, will inherently organize the system’s representation in a manner that is properly segmented and constrained in complexity at each level. As a result, while the system may still be complex, its structure becomes more intuitive and accessible to human understanding upon examination, similar to biological systems which, despite their overall intricacy, remain fundamentally comprehensible when examined at the right level of organization. \begin{itemize}
    \item \textit{Contrast:} This contrasts with the representation of the complete environment model or behavior being learned by a flat, non-decomposable model in modern neural networks trained via gradient descent, as well as the inherent assumption of having access to the entire dataset related to the full task during the learning process (see the discussion on continual learning in Section \ref{sec:ai_destructive_adaptation}).
\end{itemize}

\paragraph{Higher-Level "Regulatory" Processes:} Just as regulatory mechanisms in biology control the selective activation of developmental pathways through the targeted deployment of core processes (Section \ref{sec:edb_structure_genereg}), AI learning systems should incorporate mechanisms that can regulate the deployment of core processes at a higher level as well. Changes driven by learning should, in turn, be capable of simply adjusting “regulatory” high-level processes governing the deployment of potentially complex downstream core functions, without requiring alterations to the internal workings of these core processes—provided that the learning task can be effectively completed through such adjustments at the higher level of organization.\footnote{Note that analogues of regulatory connections from higher-level control structures have also been proposed in the brain \cite{iyer2022avoiding, hawkins2016neurons, antic2018embedded}, further reinforcing the relevance of this principle to AI.} These learned higher-order control flows and structures, over time, can become encapsulated as new core processes themselves; enabling the creation of a multi-level, hierarchical and modular structural organization of the learned model. \begin{itemize}
    \item \textit{Contrast:} This contrasts—or perhaps serves as an addition to—the ability to learn a connection solely between raw inputs and raw outputs in the black-box interpretation of neural networks. While notable fields like hierarchical reinforcement learning aim for this goal \cite{pateria2021hierarchical}, they fall far short of the versatility observed in biological systems.
\end{itemize} 

\paragraph{Growth and Local Variation \& Selection:} Learning in biological systems occurs through local variation and selection, with variation generated where necessary, enabling adaptive growth (Section \ref{sec:edb_process_varsel}). AI architectures should adopt similar local adaptation mechanisms, allowing for dynamic complexification of models without disrupting established knowledge. As discussed in Section \ref{sec:is_intelligence_evolutionary}, this approach is essential for creating new capacity—essentially a localized variation pool—where needed, ensuring that the process of learning new knowledge doesn’t interfere with existing structures. \begin{itemize}
    \item \textit{Contrast:} This contrasts with modern neural networks, which lack mechanisms for generating new structural components or sources of variation when needed for learning, without disrupting existing structures (see Section \ref{sec:contemporary_ai} for details). It's important to note that, since neural networks are already overparameterized for the task at hand—necessary for proper convergence \cite{du2018power}—this complexification-by-need is feasible, even though it may seem unscalable; since in principle, the required complexity for a given task, with proper minimal-growth or complexification methods, would be smaller than that of a typical neural network trained by SGD.
\end{itemize}

Just as the discovery and proper conceptual formulation of these mechanisms have filled the gaps in the Modern Synthesis regarding the structure of biological organisms and the progressive patterns observed in evolution, they, once transformed into principles for artificial learning systems, promise to enable continual learning and integrate a coherent structure into the learned models. Some early works in this area \cite{Erden2024Modelleyen,erden2025agential_extendedabs, erden2025agential, erden2025mnr} have already shown promising results in realizing these properties using the aforementioned design principles, we refer the reader to them for the example of a more concrete application of how they can be integrated into an AI system.

Finally, we must highlight a key insight from EDB: the complexity of core control processes indicates that the appropriate design level for adaptive systems is likely not high-level integration, but rather more proficient low-level organizational units. This stands in contrast to the majority of current ML research, which often emphasizes system architecture focusing primarily on high-level mechanisms operating on networks of artificial neurons \cite{wan2024towards, colelough2025neuro} or proposing building block alternatives with qualitatively similar capabilities \cite{bal2024rethinking}. There is limited exploration into fundamentally redesigning the capabilities of the core building blocks of ML systems. It is improbable that the design principles mentioned above can be effectively implemented across all scales of organization with fundamental entities as simple as artificial neurons or constructs of comparable complexity. Moreover, more capable computational building blocks are likely to eliminate the necessity for complex integration mechanisms layered atop the fundamental learning algorithm, as their objectives will now be achieved by the low-level units (see the handling of continual learning in \cite{erden2025agential_extendedabs,erden2025agential})—this should be viewed as a shift in the relevant design level (toward lower levels) rather than the introduction of new design goals.

\subsection{Enabling and faciliation of high-level processes with multi-level structured representations}
\label{sec:ai_future_highlevel}

One way to interpret the insights from EDB is as an explanation for the generation of the organism’s internal multi-level organizational structure.\footnote{Note that, despite some overlap, this should not be confused with multi-level selection theories \cite{damuth1988alternative, Okasha2005multilevel, gardner2015genetical, czegel2019multilevel}, which focus on evolutionary dynamics across multiple levels of organization, rather than explaining the evolution and generation of the structure of an individual multicellular organism.} Crucially, when applied to AI, this concept of a multi-level structure—characterized by hierarchy, modularity, and the correlation/reuse of common substructures—can be seen as a method for generating \textit{abstract} representations at multiple scales. These representations might include high-level percepts (such as pixels, features, and objects), equivalence sets (e.g. percepts with shared outcomes or alternative outcomes of the same percepts), or behavioral patterns (such as subpolicies in Hierarchical Reinforcement Learning \cite{bakker2004hierarchical, li2019sub}). In addition to their potential for significantly enhancing the comprehensibility of learned models, such abstract representations can enable a more separable and precise depiction of the knowledge needed for high-level cognitive processes—traditionally associated with "symbolic/classical AI"—which are notoriously difficult to integrate into modern learning systems that function as black boxes between raw observations and low-level actions/outcomes. Among these areas, the integration of learning with deliberative behavior—planning and decision-making with explicit constraints—and active information seeking stands out as particularly crucial.

As we already briefly outlined in Section \ref{sec:ai_incomprehensible}, deliberative goal-oriented behavior allows an AI agent to select actions and policies in a targeted manner, while avoiding undesired outcomes—goals or constraints that the agent either deduces in alignment with its ultimate objectives or receives externally from the designer. This is achieved without the need to re-learn the entire environment or undergo extensive re-interaction with it, as is typically required in reinforcement learning, contingent on the agent already having a functional, learned model of the environment. As detailed in Section \ref{sec:ai_incomprehensible}, there is no effective method for integrating deliberative processes with contemporary machine learning systems \cite{mcmahon2022survey, otte2015survey}, nor is there an environment-modelling approach in classical AI literature that can demonstrate a learning capability comparable to that achieved with methods like neural networks \cite{jimenez2012review, mordoch2023learning, verma2021asking, stern2017efficient}. This integration becomes feasible, however, through structured representations generated by learning—spanning abstract states, goals, and so on—allowing the identification of high-level conditions or outcomes, and linking them to abstract subgoals or subpolicies. For a demonstration of the preliminary feasibility of an approach in this spirit, see \cite{erden2025agential_extendedabs, erden2025agential}, which integrates a basic learning system based on these principles with planning capabilities.

Active information seeking naturally requires goal-oriented deliberative behavior, as the agent must deliberately act to achieve its information-seeking goals. However, this process is even further benefited by a multi-level structured representation as it also allows for a precise decomposition of uncertainty and knowledge regarding abstract (encapsulated) entities, such as percepts or behavioral patterns. This structured representation enables more focused and efficient information seeking. For instance, it allows for the explicit quantification of probabilistic relationships between observed internal variables, including their significance and reliability, facilitating directed exploration and model updates based on this quantification. This contrasts sharply with the black-box input/output representation of neural networks, where quantifying meaningful uncertainty and significance relations between identifiable percepts (raw inputs) and outcomes (predictions or behaviors) simply circles back to the objective of the overall learning system in the first place. Active information seeking is already an active area of research in neural network-based methods like reinforcement learning \cite{zhao2024active, wu2023uncertainty, mazumder2022knowledge}, yet its impact remains limited due to the challenges inherent in neural networks, as discussed throughout this article. The general approach behind these methods, however, could be applied to a structured multi-level representation, offering more precision, intuitiveness, and scalability.

It's important to note that both deliberative behavior and active information seeking are already well-established "solved problems" outside the realm of machine learning. Deliberative behavior and planning are mature fields of research that have been active since the earliest days of computing \cite{ghallab2016automated}, and information seeking is grounded in probability theory and statistical inference \cite{parr2022active, zhao2024active, wu2023uncertainty, mazumder2022knowledge} dating back centuries. There are numerous approaches in both fields that could be seamlessly integrated into learning systems once a proper structured representation is in place. These integrations would transform (learning) AI systems, moving them beyond the limitation of learning everything solely from data or experience (such as relying on reward feedback to achieve a new goal in a given environment). Instead, these systems could leverage general algorithms for deliberative behavior, reasoning, information seeking, etc. alongside specific designer knowledge for their domain (such as a high-level algorithm or entity-structure that utilizes learned low-level representations for a given environment or task), achieving the integration sought by fields like neurosymbolic AI \cite{wan2024towards,colelough2025neuro} in an organic manner that applies across all organizational levels of the system.

\subsection{Summary}

In summary, this new view of AI in a paradigm driven by these changes promises a system capable of continual learning, with the learned systems are structured and organized in a multi-level hierarchy. This structure would be comprehensible to humans upon interpretation and can be easily integrated with additional algorithms such as those for deliberative behavior, reasoning, information seeking, or domain-specific designer algorithms. The existence of such a system will undeniably enhance AI's applicability in unpredictable tasks \cite{ditzler2015learning} and those requiring greater user trust \cite{kaur2022trustworthy}. However, its potential impact extends beyond these applications. An AI system that can discover new information, seamlessly integrate this knowledge into its existing framework, and recursively expand its capabilities—much like the processes seen in evolutionary history—will profoundly influence humanity’s methods of knowledge discovery and representation, while also accelerating our technological advancement. This consideration leads us to the final major point of our discussion.



\section{Grounding Technological Singularity}
\label{sec:singularity}

No discussion of the future of AI would be complete without examining its implications, and no examination of the implications of advanced AI would be complete without discussing the concept of \textit{technological singularity}—let alone a treatment like this one, which specifically connects AI’s future to the progressive, recursively-improving nature of evolution and the mechanisms that drive it.

\subsection{Accelerating evolutionary change and singularity}
\label{sec:singularity_overview}

There are many definitions of technological singularity \cite{sandberg2013overview}, many of whom are non-exclusive and primarily emphasize different aspects of similar phenomena. We believe that despite variations in definitions, a common underlying spirit unites those that emphasize technological change and its impact on the biological aspect of the human condition \cite{kurzweil2006singularity, smart2008evo, flake2006learned, good1966speculations, vinge1993coming} (in contrast to those that focus primarily on economic or sociological aspects \cite{hanson2008economics, heylighen2007accelerating}). For all practical purposes, this spirit is captured by defining the singularity as \textit{a period of significantly accelerated technological advancements, where major breakthroughs happen within extremely short timescales, collectively driving a fundamental shift in the human condition within a timeframe that is relevant to an individual’s lifespan.}\footnote{We do not delve into the details or debates surrounding the sustainability of exponential growth or the long-term trajectory of progress—whether unbounded, saturating, or following successive S-curves. Instead, our focus is on its initiation and short-to-mid-term development before leading up to any possible saturation point.} At its core, the concept of technological singularity suggests a process in which, by its conclusion, many material problems and challenges relevant to humans will have been resolved within remarkably short periods—often understood to be as brief as days or weeks.\footnote{The term “singularity,” with its connotation of a function approaching infinity, has been a point of critique against the underlying argument. While this is a trivial point, we note that the idea of “acceleration ad infinitum” is merely a conceptual idealization, much like other uses of the singularity concept. Rather than a literal phenomenon, it signifies a critical threshold where rapid and fundamental shifts occur, similar to phase transitions \cite{johansen2001finite}.}

Rather than representing an entirely new or previously unobserved dynamic, the technological singularity should be understood as a continuation of the exponential and accelerating rise in complexity and capability of systems—alongside factors such as cognitive power \cite{kurzweil2006singularity}—that has been unfolding since the dawn of evolutionary history \cite{marc2005plausibility, smith1997major} and has been particularly evident in technological progress since at least the Industrial Revolution \cite{lucas2002industrial}. In both biological and technological evolution, the primary driver has been the ability to leverage existing core processes to generate capabilities at higher levels of organization, enabling an ever-accelerating increase in functionality. As discussed in Section \ref{sec:ms_limits}, evolutionary history demonstrates this principle. Technological evolution follows an analogous trajectory, where existing innovations are recombined into new ones \cite{fink2019mathematical, ziman2003technological, coccia2019theory}. This process interacts bidirectionally with knowledge generation, which itself operates through combinatorial mechanisms \cite{weitzman1998recombinant}: technology facilitates knowledge production, while new knowledge, in turn, drives technological development \cite{brooks1994relationship, xie2025knowledge}. We can refer to these as \textit{recursive improvement processes}.\footnote{Other factors driving developments since the Industrial Revolution, such as economic \cite{bruland2014technology} or intellectual \cite{trohler2017progressivism} drivers—which have seldom motivated large masses of people or states throughout history—can be cited. It can be argued that these are independent of recursive improvement processes; however, one could also argue that their \textit{widespread} social adoption was at least partially a consequence of the recognition of their benefits in technological development (e.g., \cite{dong2022political}). This, in turn, resulted from the increased pace of technological innovation, which led to visible technological disparities across civilizations \cite{burccak2008modernization}. Thus, these forces can be seen as products of technological advancement, at least in part, which are subsequently integrated as systemic drivers that further promote innovation.} The technological singularity, in the context of future developments, will emerge as a direct consequence of these recursive improvement processes as well.

Advanced artificial intelligence has long been seen as deeply intertwined with the concept of singularity—often as its primary driver and, in some cases, as the very foundation upon which the process is defined \cite{kurzweil2006singularity, smart2008evo, good1966speculations, vinge1993coming}. This association is far stronger than with any other prospective technology.\footnote{A possible exception to that is nanotechnology; however, even in such works (e.g. Drexler's seminal work on the topic \cite{drexler2006engines}), advanced AI is still frequently considered a prerequisite for realizing nanotechnology’s full potential.} This perspective becomes self-evident once we 
view AI not as an isolated system or merely another technological development, but rather as the autonomous coordination and regulation of computerized and cybernetic processes. In practical terms, this indicates that AI has the capability to govern the entire spectrum of our industrial, technological, and digital infrastructures—both in their current state and in future developments, provided that suitable domain-specific computational control paradigms, such as molecular or DNA computing for bioengineering or nanotechnology, are feasible.

A key point that more directly links AI to singularity—standing as the most prominent subpoint of the preceding discussion—is the role of advanced AI in autonomously driving the advancement of human knowledge (scientific discovery) and capabilities (technological development). Notably, this is also the foundation of direct recursive self-improvement mechanisms in the context of an AI-driven technological singularity \cite{lesswrongRecursiveSelfImprovement}, where an AI can hypothetically enhance itself through improvements in its algorithm, software implementation, or hardware. We will explore this point in further detail shortly.

Despite recent advancements in AI sparking discussions on its widespread applications in daily life \cite{builtinFutureChanging, ArtificialIntelligence, indiastemfoundationImpactArtificial} and even more speculative concepts like artificial general intelligence \cite{mitchell2024debates}, the idea of an AI-driven technological singularity is rarely mentioned a relevant prospect based on the current state of technology. We believe this is because there is, quite simply, no mechanistic path that can be proposed to bridge the gap between present-day AI and the recursive improvement processes necessary for singularity. Such a path would require AI systems to gradually improve their own models, continually learn from new experiences in a changing world, and do so without eroding past knowledge—all capabilities that, as discussed throughout this article, current AI systems fundamentally lack. Moreover, behavior in AI today is primarily driven by reward-based learning that demands extensive environmental interaction, rather than goal-oriented deliberation based on an internal model of the world. This, combined with the difficulty of guided information-seeking based on quantified uncertainties and reliabilities within the model itself, makes it unrealistic to expect machine learning systems to autonomously progress toward complex goals or define relevant subgoals for themselves. Whether in the pursuit of a desired real-world outcome or the strategic acquisition of new information to refine their models, current agentic AI is constrained to data-intensive reward-based learning—an approach fundamentally inadequate for driving technological singularity, as the very goal in question would, by definition, lack extensive preexisting data to train on. This is to say nothing of AI’s inability to rapidly reuse previously learned subsystems, behavioral patterns, percepts, or knowledge at a higher level to generate new processes or improve existing ones—the very essence of recursive improvement processes that drive exponential capability growth. In short, existing machine learning systems fall short \textit{qualitatively}—not merely in terms of performance metrics, but in their fundamental functional capabilities—of the prerequisites necessary to act as the prime driver of a singularity-like process.

\subsection{Why the new design paradigm for AI can fuel singularity}
\label{sec:singularity_newai}

Throughout this article, we discussed design principles that can be incorporated into AI for a new paradigm specifically aimed at resolving these particular issues (and referenced some early works that do so \cite{Erden2024Modelleyen, erden2025agential_extendedabs, erden2025agential, erden2025mnr}) that hinder AI from commencing its role in the process of technological singularity. In Section \ref{sec:ai_future_principles}, we examined the capability for continually learning without erasing past knowledge and the ability to recursively leverage previously learned processes to generate new ones. Section \ref{sec:ai_future_highlevel} addressed the prospects for integrating symbolic “non-learning” processes, such as deliberative behavior and active information acquisition. Additionally, all of these discussions were framed within the context of developments in evolutionary theory, as detailed in Section \ref{sec:edb}, which elucidates how life itself demonstrates a trend of exponentially increasing complexity and capability. Therefore, it is very reasonable to anticipate that a new paradigm of AI and artificial learning, grounded in these principles, would transcend the current barriers, enabling AI to engage in a recursive improvement process that contributes to—and further accelerates—the overall technological advancement, culminating in what conceptualize as the technological singularity.

The already-discussed aspect of this recursive improvement is the composability of high-level models, percepts, and behavior patterns from lower-level components within a structured representation (Section \ref{sec:ai_future}) to learn models and behaviors that are more complex than ever, at a faster pace than ever.\footnote{One way to rephrase this is to view singularity as a process driven by metasystem transitions, which involve creating a higher-level control system that selects between states or different exemplars of existing lower-level control systems, with the time between such transitions decreasing geometrically \cite{potapov2018technological}. In simple terms, the current AI paradigm, which relies on function approximation through large, unstructured networks, lacks any mechanism to accomodate metasystem transitions. In contrast, a new design based on the principles outlined in this article offers this possibility.} However, another crucial aspect pertains to the operational cycle an agent can embody once it acquires the capabilities discussed in the previous paragraph. More specifically, the design principles outlined here can facilitate the following cycle for AI systems:
\begin{enumerate}
    \item \textit{Improved Representation of the Environment:} A more accurate model of the environment.
    \item \textit{Improved Behavior:} With a solid model (1) in place, the agent can develop and refine behavior patterns that lead to desired changes (goals) in the environment or build upon existing behaviors to make them more efficient, effective, or adaptable to different circumstances.
    \item \textit{(Optional) Environmental Modification:} By leveraging its modeling capabilities (1) and goal-specific behavior patterns (2), an agent can alter its environment to better serve its goals (treating this as a subgoal within its overarching objectives), akin to tool-making in humans or the environmental adaptations seen in other organisms \cite{laland2016introduction}.
    \item \textit{Improved Knowledge Acquisition:} The agent’s enhanced capabilities from generating behavior patterns (2) and modifying its environment (3) boost its knowledge acquisition. A straightforward example is an agent seeking to resolve uncertainty in a distant location, needing either to access suitable transportation (2) or construct one (3). This can also be likened to the advancement of scientific capabilities through technology \cite{brooks1994relationship}.
    \item \textit{Further Improved Representation of the Environment:} This naturally follows from enhanced knowledge acquisition (4) for a system capable of integrating new information without discarding existing knowledge.
\end{enumerate}
In this cycle, each step builds upon the products of the previous steps, thus acting as a recursive improvement process for the agent's overall capabilities. The composability of higher-level model components or behavior patterns mentioned earlier is integral to points 1/5 and 2, with the cycle operating through these mechanisms as well. With this insight, it is only natural that singularity is not yet regarded as a relevant prospect in discussions of current AI technologies, given that contemporary methods entirely lack the potential for these two mechanisms of recursive improvement for AI systems.

Above, we explored the critical connection between AI's application in advancing science and technology and the prospects of technological singularity. A computerized system capable of learning new knowledge without bounds and taking appropriate actions based on a desired final goal or intermediate objectives (including information-seeking subgoals) can significantly enhance human presence across various fields, with particular benefits in scientific and technological development. Among these advantages, three particularly stand out:
\begin{itemize}
    \item AI can in principle access the entirety of human knowledge, whereas an individual can only reasonably grasp a fraction of it. This capability can rapidly accelerate the formulation of novel, otherwise-elusive scientific theories by drawing analogies from established knowledge across different fields (e.g., the Malthusian influence on Darwin’s formulation of evolution by natural selection \cite{thomson1998marginalia}). Additionally, AI can consider the interdependencies in the development of technologies that necessitate design at multiple levels of abstraction, each requiring distinct expertise and facing unique challenges.
    \item AI can address domains that are far too high-dimensional for humans to solve or represent using conventional methods. Initial applications have already emerged in areas such as quantum mechanics \cite{zheng2021artificial} and biology \cite{jumper2021highly, nobelprizeNobelPrize}, showcasing AI’s potential in tackling scientific undertakings of inherently high complexity.
    \item The knowledge acquired by an AI system is never "reset" once learned, eliminating the repeated need for costly, time-consuming, and continually extending training processes required for experts to reach a level where they can reliably advance knowledge in their fields \cite{sarrico2022expansion}.
\end{itemize}
It can be asserted that just as science and technology are regarded as the pinnacle of human intellect, they are also natural domains for applying artificial intelligence. The requirements of these fields align seamlessly with the capabilities that an advanced, capable AI should possess, such as continual learning, structrured representation, information-seeking, and goal-directedness, as discussed throughout this article. Furthermore, factors like precision, optimality, and reliability—conditioned by algorithmic foundations—of a specifically designed machine compared to what humans alone can achieve promise to significantly accelerate progress in these areas, as they do in many other applications of AI.

\subsection{Human control of accelerating technological change}
\label{sec:singularity_human_control}

Finally, we must address the control capabilities that humans can exert over artificial systems in this evolving process. Today's AI techniques, as discussed in Section \ref{sec:ai_incomprehensible}, are often marked by inherent opacity and incomprehensibility, rendering the internal representations and behaviors of such systems difficult, if not impossible, for humans to understand, design, or control. There exists an unwritten assumption that this will continue to be the case in the future. To address these challenges, research disciplines have emerged to ensure that these incomprehensible and uncontrollable systems remain “in line” \cite{ji2023ai}, aiming to mitigate potential negative consequences. This perspective is reflected in discussions about the prospects of singularity, where many analyses, either implicitly or explicitly, portray humans as passive observers rather than active participants in the process once advanced AI capable of autonomously driving recursive improvement is developed.

A perspective on AI that emphasizes and prioritizes structural organization, like the one outlined here, has the potential to change this game. To repeat our point in Section \ref{sec:ai_future_principles}, structural organization—particularly a modular and hierarchical composition that develops during the learning process—can naturally arrange the system’s representation in a way that is appropriately decoupled and limited in complexity at each level, leading to a more intuitive structure that could be comprehensible to humans upon inspection. In this framework, the internal representation of an AI system would differ significantly from the billions of continuous-valued parameters found in densely connected networks of current machine learning systems. Instead, it would resemble object-oriented software organized hierarchically, a modular multi-scale device, or a biological organism. While it is true that such a representation could still be vastly complex—requiring multiple experts with diverse specialties to fully decipher the entire system—this is also true of complex software, devices, or biological organisms, which are generally regarded as fundamentally comprehensible and modifiable by human intellect conditioned on expertise and methodological examination. There is no compelling reason to believe that incomprehensibility and uncontrollability (as it pertains to the direct ability of human intellect to modify representations or behaviors, rather than simply bias behavior toward a desired outcome through external pressures during training or inference) must be inherent features of AI systems. Although it is certainly possible for such an AI system to enter a recursive improvement cycle and make progress without human supervision when employing structured representations, we contend that the nature of the process —whether it operates independently of human oversight, or is closely controlled/guided by humans, or exists somewhere in between— will be more of a choice rather than an inevitability.

\subsection{Summary}

Major treatments of the singularity to date, particularly regarding the role of advanced AI, have often left key points ambiguous, giving the concept an air of speculation or even mysticism due to their grand claims lacking specificity. Our goal was to dispel this impression by firmly linking the idea of exponentially increasing progression to both the well-documented history of evolution—an area that is widely studied and far less speculative—and the specific requirements and mechanisms that AI must possess to achieve this progression (as conceptually discussed throughout the text, with references to early specific works noted—see \cite{Erden2024Modelleyen, erden2025agential_extendedabs, erden2025agential, erden2025mnr}). Specifically, we examined how learning without the destruction of past knowledge, a structured internal representation, and mechanisms for deliberation \& active information-seeking can create a combination that enables unbounded, rapid learning and modification of the external world in a way that fuels further learning and goal realization with an expanding the repertoire of previously acquired knowledge and capabilities. We drew parallels with newly formulated principles in evolutionary theory that elucidate how exponential increases in complexity can arise from a recursive improvement process that leverages previously evolved processes to spur further evolution. We argue that this intellectual framework portrays an accelerating technological progression driven by advanced AI capable of recursive improvement (including recursive \textit{self}-improvement), framed as a technological singularity, as something conceptually grounded and non-speculative. Furthermore, we addressed another common theme in discussions of the singularity: the notion that the process doomed to be uncontrollable and independent of human influence. We clarified that this does not have to be the case with structured internal representations generated by systems designed according to the principles outlined in this article, alleviating some anxieties surrounding this prospect and allowing us to focus on creating AI systems that can be properly understood and effectively controlled by humans.

Although these prospects may appear distant and have largely been absent from mainstream discourse, we believe they might be closer than they seem. The conceptual foundation for this possibility, as outlined in this article, already exists, supported also by early demonstrative works. With the enhancement or complete redesign of AI methods based on the framework discussed here, the prospect of an autonomous, recursively accelerating paradigm in scientific and technological advancement—driven by AI capable of unbounded learning, integrating structured representations, and performing non-learning tasks such as deliberative behavior and active knowledge-seeking in an integrated manner—represents a very tangible possibility for the near future.

\section{Conclusion}

The design paradigm of contemporary AI, despite their widespread adoption and remarkable success in areas previously deemed unsolvable by "classical" techniques, have inherent limitations regarding essential qualitative capabilities expected from any effective learning system. These limitations hinder AI’s potential, particularly their inability to learn new knowledge without destroying previously acquired knowledge. Additionally, the unstructured, intertwined, and overparameterized internal representations generated by current machine learning approaches obstruct comprehensibility and integration with non-learning processes, such as deliberation and explicit uncertainty representation as crucial for active information acquisition.

The current approach to AI, with its strengths, limitations, and underlying assumptions, is analogous to the Modern Synthesis view of evolution, the dominant perspective of the 20\textsuperscript{th} century. Recent advancements in evolutionary theory, especially in the field of Evolutionary Developmental Biology, have successfully addressed the shortcomings of the Modern Synthesis, providing a more comprehensive and intuitive understanding of life’s evolution on Earth, as well as better explanations for observed patterns in evolutionary history. The principles of EDB, when translated into design principles for AI, hold the potential to initiate a new design paradigm that effectively overcomes the limitations of existing systems. This will have significant downstream implications, particularly the potential for AI to fuel a process that can be conceptualized as a technological singularity, as the barriers to this phenomenon have largely arisen from the aforementioned limitations of AI, rendering the prospects for such a development both grounded and plausible.

\printbibliography

\end{document}